\input ./style/arxiv-general.cfg
\documentclass[aoas,MSNbibl,nameyear,dvips]{arximspdf}
\makeatletter
   \@ifpackageloaded{graphicx}{}{\usepackage{graphicx}}
\makeatother
\usepackage{url,breakurl}

\doi{10.1214/15-AOAS862}
\volume{9}
\issue{4}
\pubyear{2015}
\firstpage{1761}
\lastpage{1791}
\docsubty{FLA}

\makeatletter
\let\epsilon\varepsilon
\newcommand{\rrVert}{\Vert}
\newcommand{\llVert}{\Vert}
\newtheorem{prop}{Proposition}
\newcommand{\ind}{\stackrel{\mathrm{ind}}{\sim}}
\newcommand{\bu}{\mathbf{u}}
\newcommand{\bz}{\mathbf{z}}
\newcommand{\by}{\mathbf{y}}
\newcommand{\bfbeta}{\bolds{\beta}}
\makeatother

\begin{document}
\begin{frontmatter}

\title{Multivariate spatio-temporal models for high-dimensional
areal data with application to~Longitudinal Employer-Household Dynamics\thanksref{T1}}
\runtitle{Multivariate spatio-temporal models}

\begin{aug}
\author[A]{\fnms{Jonathan R.}~\snm{Bradley}\corref{}\ead[label=e1]{bradleyjr@missouri.edu}},
\author[A]{\fnms{Scott H.}~\snm{Holan}}\\
\and
\author[A]{\fnms{Christopher K.}~\snm{Wikle}}
\runauthor{J. R. Bradley, S. H. Holan and C. K. Wikle}
\affiliation{University of Missouri}
\address[A]{Department of Statistics\\
University of Missouri\\
146 Middlebush Hall\\
Columbia, Missouri 65211\\
USA\\
\printead{e1}}
\end{aug}
\thankstext{T1}{Supported in   part by
the U.S. National Science Foundation (NSF) and the U.S. Census Bureau
under NSF Grant SES-1132031, funded through the NSF-Census Research
Network (NCRN) program.}

%
\received{\smonth{1} \syear{2015}}
%
\revised{\smonth{6} \syear{2015}}

\begin{abstract}
Many data sources report related variables of interest that are {also}
referenced over geographic regions and time; however, there are
relatively few general statistical methods that one can readily use
that incorporate these multivariate spatio-temporal dependencies.
Additionally, many multivariate spatio-temporal areal data sets are
extremely high dimensional, which leads to practical issues when
formulating statistical models. For example, we analyze Quarterly
Workforce Indicators (QWI) published by the US Census Bureau's
Longitudinal Employer-Household Dynamics (LEHD) program. QWIs are
available by different variables, regions, and time points, resulting
in millions of tabulations. Despite their already expansive coverage,
by adopting a fully Bayesian framework, the scope of the QWIs can be
extended to provide estimates of missing values along with associated
measures of uncertainty. Motivated by the LEHD, and other applications
in federal statistics, we introduce the multivariate spatio-temporal
mixed effects model (MSTM), which can be used to efficiently model
high-dimensional multivariate spatio-temporal areal data sets. The
proposed MSTM extends the notion of Moran's I basis functions to the
multivariate spatio-temporal setting. This extension leads to several
methodological contributions, including extremely effective dimension
reduction, a dynamic linear model for multivariate spatio-temporal
areal processes, and the reduction of a high-dimensional parameter
space using {a novel} parameter model.
\end{abstract}

\end{frontmatter}

\section{Introduction}\label{secintro}

Ongoing data collection from the private sector along with federal,
state, and local governments have produced massive quantities of data
measured over geographic regions (areal data) and time. This
unprecedented volume of spatio-temporal data contains a wide range of
variables and, thus, has created unique challenges and opportunities
for those practitioners seeking to capitalize on their full utility.
For example, methodological issues arise because these data exhibit
complex multivariate spatio-temporal covariances that may involve
nonstationarity and interactions between different\vadjust{\goodbreak} variables, regions,
and times. Additionally, the fact that these data (with complex
dependencies) are often extremely high-dimensional (so called ``big
data'') leads to the important practical issue associated with computation.

As an example, the US Census Bureau's Longitudinal Employer-Household
Dynamics (LEHD) program produces estimates of US labor force variables
called Quarterly Workforce Indicators (QWIs). The QWIs are derived from
a combination of administrative records and data from federal and state
agencies [\citet{abowd}]. The sheer amount of QWIs available is
unprecedented, and has made it possible to investigate local (in
space--time) dynamics of several variables important to the US economy.
For example, the average monthly income QWI is estimated quarterly over
multiple regions and industries (e.g., education, manufacturing, etc.).
In total, there are 7{,}530{,}037 quarterly estimates of average monthly income.

The QWIs present interesting methodological challenges. In particular,
not every state signs a new Memorandum of Understanding (MOU) each year
and, hence, QWIs are not provided for these states [\citet{abowd}, Section~5.5.1].
Furthermore, some data are suppressed at certain regions
and time points due to disclosure limitations [\citet{abowd}, Section~5.6]. Another limitation is that uncertainty measures are not
made publicly available. Consequently, it is difficult for QWI data
users to assess the quality of the published estimates. Thus, producing
a complete set of estimates (i.e., national coverage) that have
associated measures of uncertainty is extremely important and provides
an unprecedented tool for the LEHD-user community.
As such, we take a fully Bayesian approach to estimating quarterly
measures of average monthly income and, thus, provide a complete set of
estimates that have associated measures of uncertainty.

A fully Bayesian model that can efficiently and jointly model a
correlated (over multiple variables, regions, and times) data set of
this size ($7.5 \times 10^{6}$) is unprecedented. It is instructive to
compare the dimensionality of the QWI to data sets used in spatial
analyses in other scientific domains. For example, \citet{banerjee} use
a fully Bayesian approach to analyze a multivariate spatial
agricultural data set consisting of 40{,}500 observations; \citet{johan}
use an empirical Bayesian approach to analyze a spatial data set of
total column ozone with 173{,}405 observations; \citet{lindgren-2011} use
a fully Bayesian approach to analyze climate spatially using
approximately 32{,}000 observations; and \citet{aritrajsm} use an
empirical Bayesian approach to analyze cloud fractions using a data set
of size 2{,}748{,}620.
Furthermore, none of these methods allow for
multivariate dependencies between different geographic regions and time points.

Despite the wide availability of high-dimensional areal data sets
exhibiting multivariate spatio-temporal dependencies, the literature on
modeling multivariate spatio-temporal areal processes is relatively
recent by comparison. For example, various multivariate space--time
conditional autoregressive (CAR) models have been proposed by
\citet{carlinmst},\vadjust{\goodbreak} \citet{congdon}, \citet{pettitt}, \citet{zhuglm},
\citet{daniels}, and \citet{bestmst}, among others. However,\vadjust{\goodbreak} these
methodologies cannot efficiently model high-dimensional data sets.
Additionally, these approaches impose separability and various
independence assumptions, which are not appropriate for many settings,
as these models fail to capture important interactions and dependencies
between different variables, regions, and times [\citet{steinSep}].
Hence, we introduce the multivariate spatio-temporal mixed effects
model (MSTM) to analyze high-dimensional multivariate data sets that
vary over different geographic regions and time points.

The MSTM is built upon the first order linear dynamic spatio-temporal
model (DSTM) [\citet{cressie-wikle-book}]. To date, no DSTM has been
proposed to analyze multivariate high-dimensional areal data, and, as a
result, the components of the MSTM require significant methodological
development. Specifically, we introduce novel classes of multivariate
spatio-temporal basis functions, propagator matrices, and parameter
models to be used within the MSTM.

The components of the MSTM can be specified to have a computationally
advantageous reduced rank structure [e.g., see \citet{wikleHandbook}],
which allows us to analyze high-dimensional areal data (e.g., QWIs from
the LEHD program). This reduced rank structure is achieved, in part, by
extending various aspects of the model suggested by \citet{hughes} from
the univariate spatial-only setting to the multivariate spatio-temporal
setting. Specifically, we extend the Moran's I (MI) basis functions to
the multivariate spatio-temporal setting
[for the spatial-only case see \citeauthor{griffith2000} (\citeyear{griffith2000},
\citeyear{griffith2002}, \citeyear{griffith2004}), \citet{griffith2007}, \citet{hughes}, \citet{aaronp}]. Further, we propose a novel propagator (or
transition) matrix for the first-order vector autoregressive---VAR(1)---model, which we call the MI propagator matrix. In this context, the
propagator matrix of {the} VAR(1) model is specified to have a
desirable nonconfounding property, which is similar to the
specification of the multivariate spatio-temporal MI basis functions.

We also propose an extension of the spatial random effects covariance
parameter model used in \citet{hughes} and \citet{aaronpBayes}, which
we call the MI prior. Here, we interpret the MI prior as a rescaling of
the covariance matrix that is specified to be close (in Frobenius norm)
to a ``target precision'' matrix. This parameterization significantly
reduces the dimensionality of the parameter space, thereby reducing the
computational burden associated with fully Bayesian inference in
high-dimensional spatio-temporal settings. Furthermore, this target
precision matrix can be sensibly chosen based on knowledge of the
underlying spatial process.

In addition to modeling QWIs from the LEHD, the MSTM can be used to
effectively address numerous statistical modeling and analysis problems
in the context of multivariate spatio-temporal areal data. For example,
besides analyzing high-dimensional data, the MSTM can also be used to
model nonseparable and nonstationary covariances, and to combine data
from multiple repeated surveys. Although we mainly focus on modeling
high-dimensional\vadjust{\goodbreak} multivariate spatio-temporal areal data (e.g., QWIs
from the LEHD), the MSTM is tremendously flexible and can be readily
adapted to other settings.

The remainder of this article is organized as follows. In Section~\ref{secmotiv} we introduce the LEHD-QWI data set and further describe the
methodological challenges that we consider. Next, in Section~\ref{secMSTM} we provide mathematical foundations for the MSTM. Then, in
Section~\ref{secspecs} we introduce the multivariate spatio-temporal
MI basis functions, the MI propagator matrix, and the parameter model
for the covariance matrix of the random effects term. Section~\ref{analysis} provides an empirical study that is used to evaluate the
effectiveness of the MSTM in recovering the unobserved latent process
(``true'' underlying values). Additionally, in Section~\ref{analysis}
we use the MSTM to jointly analyze all 7{,}530{,}037 QWIs obtained from the
US Census Bureau's LEHD program. Finally, Section~\ref{secdisc}
contains discussion. For convenience of exposition, proofs of the
technical results and details surrounding the MCMC algorithm are left
to an \hyperref[app]{Appendix}.

\section{LEHD---Quarterly Workforce Indicators}\label{secmotiv}
\begin{figure}

\includegraphics{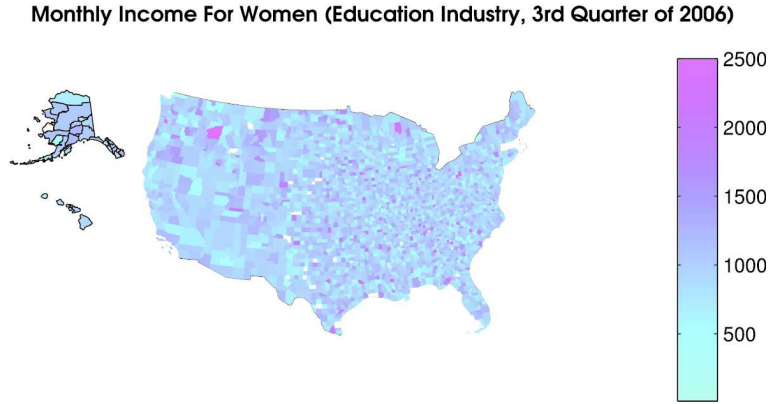}

\caption{We present the QWI for quarterly average
monthly income (US dollars) for 2970 US counties, for women, for the
education industry, and for the third quarter of 2006. The white areas
indicate QWIs that are not made available by LEHD.}
\label{figobsmap}
\end{figure}

The
LEHD program provides public access QWIs on several earnings variables
for each quarter of the year over various geographies of the US (\surl{http://www.census.gov/}). For a comprehensive
description regarding the creation of QWIs, see \citet{abowd}. Here, we
consider quarterly measures of average monthly income for individuals
with steady jobs. A subset of this data set representing QWIs for 2970
US counties for women in the education industry during the third
quarter of 2006 is displayed in Figure~\ref{figobsmap}. However, the
QWIs are much more extensive. Specifically, the quarterly average
monthly income for individuals that have a steady job is available over
92 quarters (ranging from 1990 to 2013), all of the 3145 US counties,
by each gender, and by {20 different industries}. This results in the
aforementioned data set having 7{,}530{,}037 observations---which we model jointly.

The high-dimensional nature of QWIs and expansive coverage (e.g., quarterly average monthly incomes) allows economists and other
subject matter researchers to study differences in key US economic
variables over many regions and times. Consequently, QWIs have had a
significant impact on the economics literature; for example, see \citet
{supermarket}, \citet{thompson}, \citet{workpaper}, \citet{irle}, among
others. This demand for QWIs shows a clear need for developing
statistical methodology that can be used to analyze such
high-dimensional data sets. The current statistical approaches
available cannot capitalize on the full utility of the QWIs. For
example, \citet{abowdlmm} limit the spatial and temporal scope of their
analysis, which allows them to efficiently analyze only a portion of
the QWIs.

The complexity of the QWIs is further exacerbated by missing values; by
``missing'' we mean that the QWI is not provided by the LEHD program.
Consider the (quarterly) average monthly income example, the total
gender/industry/space/time combinations results in $2\times20\times
3145 \times92 = 11{,}573{,}600$ possible QWIs. Hence, roughly $35\%$ of
the QWIs are missing. This leads to a total of $11{,}573{,}600^{2}$
pairwise covariances that require modeling using random effects.
Nevertheless, allowing for multivariate spatio-temporal covariances is
extremely important from the perspective of predicting (imputing)
missing QWIs.

As an example, in Figure~\ref{figobsmap}, one might expect the
quarterly average monthly income for men to be associated with the
value for quarterly average monthly income for women. Likewise, nearby
observations in space and time are often similar in value
[\citet{cressie-wikle-book}]. If no multivariate spatio-temporal dependencies
are present in the data, then one can not borrow strength among
``similar'' variables and ``nearby'' observations to improve the
precision of the estimated QWIs. An exploratory analysis, based on the
empirical covariance matrices computed from the log QWIs (not shown),
indicates that the QWIs are indeed correlated across different
variables, regions, and times. Consequently, this suggests that a
statistical model that allows for multivariate spatio-temporal
dependence can be efficiently utilized to predict (impute) QWIs.


\section{The multivariate spatio-temporal mixed effects model}\label{secMSTM}
The DSTM framework is a well-established modeling approach used to
analyze data referenced over space and time. This approach is extremely
flexible since it allows one to define how a group of spatial regions
temporally evolve [e.g., see  \citet{cressie-wikle-book}, page~13], as
opposed to defining the temporal evolution of a process at each
geographic region of interest. The MSTM represents a novel extension of
the DSTM to the multivariate areal data setting, where we now allow
groups of spatially referenced variables to evolve over time. Thus, in
Sections~\ref{sec31} and \ref{secprocess} we introduce the MSTM in terms of the familiar
``data model'' and ``process model'' DSTM terminology [\citet{cressie-wikle-book}].

\subsection{The MSTM data model}\label{sec31}
The data model for the MSTM is
defined as
%
\begin{eqnarray}
Z_{t}^{(\ell)}(A) &=&  Y_{t}^{(\ell)}(A)
+ \epsilon_{t}^{(\ell
)}(A);
\nonumber
\\[-8pt]
\label{datamodel}
\\[-8pt]
\eqntext{\displaystyle
 \ell= 1,\ldots,L, t =
T_{L}^{(\ell)},\ldots ,T_{U}^{(\ell)}, A \in
D_{\mathrm{P},t}^{(\ell)},}
\end{eqnarray}
where $\{Z_{t}^{(\ell)}(\cdot)\}$ represents multivariate
spatio-temporal areal data. The components of (\ref{datamodel}) are
defined and elaborated as follows:
\begin{enumerate}[4.]
\item[1.] The subscript ``$t$'' denotes discrete time, and the superscript
``$\ell$'' indexes different variables of interest (e.g., the QWI for
women in the education industry). There are a total of $L$ variables of
interest (i.e., $\ell= 1,\ldots,L$) and we allow for a different
number of observed\vspace*{1pt} time points for each of the $L$ variables of
interest (i.e., for variable $\ell$, $t = T_{L}^{(\ell)},\ldots
,T_{U}^{(\ell)}$).
\item[2.] We require $T_{L}^{(\ell)},\ldots,T_{U}^{(\ell)}$ to be on the
same temporal scale (e.g., quarterly) for each $\ell$, $T_{L}^{(\ell)}
\le T_{U}^{(\ell)}$, $\min (T_{L}^{(\ell)} ) =
1$, and $\max  (T_{U}^{(\ell)} ) = T \ge1$.
\item[3.] The set $A$ represents a generic areal unit. For example, a given
set $A$ might represent a state, county, or a census tract. Denote the
collection of all $n_{t}^{(\ell)}$ observed areal units with the set
$D_{\mathrm{O},t}^{(\ell)}\equiv\{A_{t,i}^{(\ell)}: i = 1,\ldots
,n_{t}^{(\ell)}\}$; $\ell= 1,\ldots,L$. The observed data locations
are different from the prediction locations $D_{\mathrm{P},t}^{(\ell
)}\equiv\{A_{t,j}^{(\ell)}: j = 1,\ldots,N_{t}^{(\ell)}\}$, that is,
we consider predicting on a spatial support that may be different from
$\{D_{\mathrm{O},t}^{(\ell)}\}$ (e.g., the counties with missing QWIs
are not included in $\{D_{\mathrm{O},t}^{(\ell)}\}$, but are included
in $\{D_{\mathrm{P},t}^{(\ell)}\}$). Additionally, denote the number of
prediction locations at time $t$ as $N_{t} =\sum_{\ell=
1}^{L}N_{t}^{(\ell)}$ and the total number of prediction locations as
$N \equiv\sum_{t = 1}^{T}N_{t}$. In a similar manner, the number of
observed locations at time $t$ and total number of observations are
given by $n_{t} =\sum_{\ell= 1}^{L}n_{t}^{(\ell)}$ and $n \equiv
\sum_{t = 1}^{T}n_{t}$, respectively.
\item[4.] The random process $Y_{t}^{(\ell)}(\cdot)$ represents the $\ell
$th variable of interest at time~$t$. For example, $Y_{t}^{(\ell
)}(\cdot
)$ might represent the quarterly average monthly income for women in
the education industry at time~$t$. The stochastic properties of $\{
Y_{t}^{(\ell)}(\cdot)\}$ are defined in Section~\ref{secprocess}.
Latent processes like $\{Y_{t}^{(\ell)}(\cdot)\}$ have been used to
incorporate spatio-temporal dependencies [e.g., see \citet{cressie-wikle-book}],
which we modify to the multivariate
spatio-temporal areal data setting.

\item[5.] It is assumed that $\epsilon_{t}^{(\ell)}(\cdot)$ is a
white-noise Gaussian process with mean zero and unknown variance $\operatorname{var}\{
\epsilon_{t}^{(\ell)}(\cdot)\}= v_{t}^{(\ell)}(\cdot)$ for $\ell=
1,\ldots,L$, and $t = T_{L}^{(\ell)},\ldots,T_{U}^{(\ell)}$. The
presence of $\{\epsilon_{t}^{(\ell)}(\cdot)\}$ in (\ref{datamodel})
allows us to take into account that we do not perfectly observe $\{
Y_{t}^{(\ell)}(\cdot)\}$, and instead observe a {noisy} version $\{
Z_{t}^{(\ell)}(\cdot)\}$. In many settings, there is information that
we can use to define $\{\epsilon_{t}^{(\ell)}(\cdot)\}$ (e.g.,
information provided by the statistical agency). If one does not
account for this extra source of variability, then the total
variability of the process $\{Y_{t}^{(\ell)}(\cdot)\}$ may be
underestimated. For example, \citet{finley} show that if one ignores
white-noise error in a Gaussian linear model, then one underestimates
the total variability of the latent process of interest.
\end{enumerate}

\subsection{The MSTM process model}\label{secprocess}
The process model for MSTM is defined as
%
\begin{eqnarray}
&& Y_{t}^{(\ell)}(A) = \mu_{t}^{(\ell)}(A)
+ \mathbf{S}_{t}^{(\ell
)}(A)^{\prime}\bolds{
\eta}_{t} + \xi_{t}^{(\ell)}(A);
\nonumber
\\[-8pt]
\label{processmodel}
\\[-8pt]
\eqntext{\displaystyle
 \ell= 1,
\ldots,L, t = T_{L}^{(\ell)},\ldots,T_{U}^{(\ell)},A
\in D_{\mathrm
{P},t}^{(\ell)}.}
\end{eqnarray}
In (\ref{processmodel}), $Y_{t}^{(\ell)}(\cdot)$ represents the
$\ell
$th spatial random process of interest at time $t$, which is modeled by
three terms on the right-hand side of (\ref{processmodel}). The first
term [i.e., $\{\mu_{t}^{(\ell)}(\cdot)\}$] is a fixed effect, which is
unknown, and requires estimation. We set $\mu_{t}^{(\ell)}(\cdot
)\equiv
\mathbf{x}_{t}^{(\ell)}(\cdot)^{\prime}\bolds{\beta}_{t}$, where
$\mathbf{x}_{t}^{(\ell)}$ is a known $p$-dimensional vector of covariates and
$\bolds{\beta}_{t}\in\mathbb{R}^{p}$ is an associated unknown parameter
vector; $\ell= 1,\ldots,L$ and $t = 1,\ldots,T$. In general, we allow
both $\mathbf{x}_{t}^{(\ell)}$ and $\bolds{\beta}_{t}$ to change over
time; however, in practice, one must assess whether or not this is
appropriate for a given application. For the QWI example we specify
$\mathbf{x}_{t}^{(\ell)}$ and $\bolds{\beta}_{t}$ to be constant over time.

The second term on the right-hand side of (\ref{processmodel}) [i.e.,
$\{\mathbf{S}_{t}^{(\ell)}(\cdot)^{\prime}\bolds{\eta}_{t}\}$] represents
multivariate spatio-temporal dependencies. The $r$-dimensional vectors
of multivariate spatio-temporal basis functions $\mathbf{S}_{t}^{(\ell
)}(\cdot)\equiv(S_{t,1}^{(\ell)}(\cdot),\ldots,S_{t,r}^{(\ell
)}(\cdot
))^{\prime}$ are prespecified for each $t = 1,\ldots,T$ and $\ell=
1,\ldots,L$, and in Section~\ref{secmibf} we propose a new class of
multivariate spatio-temporal basis functions to use in (\ref
{processmodel}). The $r$-dimensional random vector $\bolds{\eta}_{t}$ is
assumed to follow a spatio-temporal VAR(1) model  [\citet{cressie-wikle-book}, Chapter~7]
%
\begin{equation}
\label{autoregressive} \bolds{\eta}_{t} = \mathbf{M}_{t}\bolds{
\eta}_{t-1} + \bu_{t};\qquad t = 2,3,\ldots,T,
\end{equation}
where for all $t$ the $r$-dimensional random vector $\bolds{\eta}_{t}$ is
Gaussian with mean zero and has an unknown $r\times r$ covariance
matrix $\mathbf{K}_{t}$; $\mathbf{M}_{t}$ is a $r \times r$ known
propagator matrix (see discussion below); and $\bu_{t}$ is an
$r$-dimensional Gaussian random vector with mean zero and unknown
$r\times r$ covariance matrix $\mathbf{W}_{t}$ and is independent of
$\bolds{\eta}_{t-1}$.

First order vector autoregressive models may offer more realistic
structure with regards to interactions across space and time. This is a
feature that cannot be included in the alternative modeling approaches
discussed in Section~\ref{secintro}. Additionally, the (temporal)
VAR(1) model has been shown to perform well (empirically) in terms of
both estimation and prediction for federal data repeated over time
[\citet{Jones}, \citet{Bell}, \citet{Feder}].

The $r$-dimensional random vectors $\{\bolds{\eta}_{t}\}$ are not only
used to model temporal dependencies in $\{Y_{t}^{(\ell)}(\cdot)\}$, but
are also used to model multivariate dependencies. Notice that the
random effect term $\bolds{\eta}_{t}$ is common across all $L$ processes.
Allowing for a common random effect term between different processes is
a straightforward way to induce dependence [\citet{cressie-wikle-book}, Chapter~7.4].
This strategy has been previously used in the
univariate spatial and multivariate spatial settings [e.g., see
\citet{royle1999}, \citet{finley}, and \citet{finley2}] and has been
extended here.

Finally, the third term on the right-hand side of (\ref{processmodel})
[i.e., $\{\xi_{t}^{(\ell)}(\cdot)\}$] represents fine-scale variability
and is assumed to be Gaussian white noise with mean zero and unknown
variance $\{\sigma_{\xi,t}^2\}$. In general, $\{\xi_{t}^{(\ell
)}(\cdot
)\}$ represents the leftover variability not accounted for by $\{
\mathbf{S}_{t}^{(\ell)}(\cdot)^{\prime}\bolds{\eta}_{t}\}$. One might consider
modeling spatial covariances in $\{\xi_{t}^{(\ell)}(\cdot)\}$. Minor
adjustments to our methodology could be used to incorporate, for
example, a CAR model [\citet{banerjee-etal-2004}, Chapter~3], tapered
covariances [\citet{cressie}, page~108], or block diagonal covariances
[\citet{steinr}] in $\{\xi_{t}^{(\ell)}(\cdot)\}$.

\section{Multivariate spatio-temporal mixed effects model
specifications} \label{secspecs}
Many specifications of the MSTM require methodological development
before one directly can apply it to the QWIs. In particular, we need to
specify the multivariate spatio-temporal basis functions $\{\mathbf
{S}_{t}^{\ell}(\cdot)\}$, the propagator matrices $\{\mathbf{M}_{t}\}$,
and the parameter models for $\{\mathbf{K}_{t}\}$ and $\{\mathbf
{W}_{t}\}$. These contributions are detailed in Sections~\ref{secmibf}, \ref{secmipr}, and \ref{secparmod}, respectively.

\subsection{Moran's I basis functions} \label{secmibf}
In
principle,  the $r$-dimensional vector $\mathbf{S}_{t}^{(\ell)}(\cdot)$
can belong to any class of spatial basis functions; however, we use
Moran's I (MI) basis functions, since they have many properties that
are needed to accurately and efficiently model QWIs. In particular, the
MI basis functions can be used to model areal data in a reduced
dimensional space (i.e., $r \ll n$). This feature allows for fast
computation of the distribution of $\{\bolds{\eta}_{t}\}$, which can
become computationally expensive for large $r$. This will be especially
useful for analyzing the QWIs in Section~\ref{secmassive}, which
{consists of 7{,}530{,}037 observations}. Additionally, the MI basis
functions allow for nonstationarity in space, which is a realistic
property for modeling QWIs (see Section~\ref{secmotiv} for a discussion).

A defining (and mathematically desirable) property of the MI basis
functions is that they guarantee there are no issues with confounding
between fixed and random effects. {This property of removing any
confounding frees us to consider inferential questions in addition to
multivariate spatio-temporal prediction. For example, the QWIs can be
used to investigate the degree of gender inequality in the US by
comparing the mean (i.e., $\mu_{t}^{(\ell)}$) average monthly income
for men and women, respectively.}

Thus, to derive MI basis functions to use for QWIs, we extend this
defining property to the multivariate spatio-temporal setting. Here,
the derivation starts with the MI operator. Recall that the MI
statistic is a measure of association, which equals to a weighted sums
of squares where the weights are called the MI operator [see \citet{hughes}].
At time $t$ the MI operator is explicitly defined as
%
\begin{eqnarray}
&& \mathbf{G}(\mathbf{X}_{t},\mathbf{A}_{t})
\equiv \bigl(\mathbf {I}_{N_{t}} - \mathbf{X}_{t} \bigl(
\mathbf{X}_{t}^{\prime}\mathbf {X}_{t}
\bigr)^{-1}\mathbf{X}_{t}^{\prime} \bigr)\mathbf
{A}_{t} \bigl(\mathbf{I}_{N_{t}} - \mathbf{X}_{t}
\bigl(\mathbf{X}_{t}^{\prime
}\mathbf {X}_{t}
\bigr)^{-1}\mathbf{X}_{t}^{\prime} \bigr);
\nonumber
\\[-8pt]
\label{mioperator}
\\[-8pt]
\eqntext{\displaystyle t = 1,
\ldots,T,}
\end{eqnarray}
where the $N_{t}\times p$ matrix $\mathbf{X}_{t} \equiv (\mathbf
{x}_{t}^{(\ell)}(A): \ell= 1,\ldots,L, A \in D_{\mathrm
{P},t}^{(\ell
)} )^{\prime}$, $\mathbf{I}_{N_{t}}$ is an $N_{t}\times N_{t}$
identity matrix, and $\mathbf{A}_{t}$ is the $N_{t}\times N_{t}$
adjacency matrix corresponding to the edges formed by $\{D_{\mathrm
{P},t}^{(\ell)}:\ell= 1,\ldots,L\}$. Notice that the MI operator in~(\ref{mioperator}) defines a column space that is orthogonal to
$\mathbf
{X}_{t}$. This can be used to ensure nonconfounding between $\bolds{\beta
}_{t}$ and $\bolds{\eta}_{t}$. Specifically, from the spectral
representation $\mathbf{G}(\mathbf{X}_{t},\mathbf{A}_{t}) = \bolds{\Phi}_{X,G,t}\bolds{\Lambda}_{X,G,t}{\bolds{\Phi}_{X,G,t}^{\prime}}$, we denote
the $N_{t}\times r$ real matrix formed from the first $r$ columns of
$\bolds{\Phi}_{X,G,t}$ as $\mathbf{S}_{X,t}$. Additionally, we set the row
of $\mathbf{S}_{X,t}$ that corresponds to variable $\ell$ and areal
unit $A$ equal to $\mathbf{S}_{t}^{(\ell)}(A)$. Thus, by definition,
for each $t$ the $N_{t}\times p$ matrix of covariates $\mathbf{X}_{t}$
is linearly independent of the columns of the $N_{t}\times r$ matrix of
basis functions $\mathbf{S}_{X,t}$ and, hence, there are no issues with
confounding between $\bolds{\beta}_{t}$ and $\bolds{\eta}_{t}$.

{It is important to emphasize that the orthogonalization of $\mathbf
{X}_{t}$ to obtain $\mathbf{S}_{X,t}$ is done over the support of the
entire spatial region (i.e., $D_{\mathrm{P},t}$), which removes
confounded random effects at any prediction location of interest. In
principal, one might use an orthogonalization over a subset, say $D
\subset D_{\mathrm{P},t}$, and use a different class of basis functions
to define $\mathbf{S}_{t}^{(\ell)}(\cdot)$ at prediction locations
outside $D$. However, in this case prediction locations outside $D$ may
suffer from problems with confounding and, hence, inference on the
underlying mean $\mu_{t}^{(\ell)}(\cdot)$ may be incorrect.}

\subsection{Moran's I propagator matrix} \label{secmipr}
The problem
of confounding provides motivation for the definition of the MI basis
functions $\{\mathbf{S}_{X,t}^{(\ell)}(\cdot)\}$. In a similar manner,
the problem of confounding manifests in a spatio-temporal VAR(1) model
and can be addressed through careful specification of $\{\mathbf
{M}_{t}\}$. To see this, substitute (\ref{autoregressive}) into~(\ref
{processmodel}) to obtain
%
\begin{equation}
\label{matrixprocess} \by_{t} = \mathbf{X}_{t}\bolds{
\beta}_{t} + \mathbf{S}_{X,t}\mathbf {M}_{t}\bolds{
\eta}_{t-1} + \mathbf{S}_{X,t}\bu_{t} + \bolds{\xi
}_{t};\qquad t = 2,\ldots, T,
\end{equation}
where $\by_{t}\equiv(Y_{t}^{(\ell)}(A): \ell= 1,\ldots,L, A \in
D_{\mathrm{P},t}^{(\ell)})^{\prime}$ and $\bolds{\xi}_{t}\equiv(\xi
_{t}^{(\ell)}(A): \ell= 1,\ldots,L$, $A \in D_{\mathrm{P},t}^{(\ell
)})^{\prime}$ are $N_{t}$-dimensional latent random vectors. The
specification of $\{\mathbf{S}_{X,t}\}$ using MI basis functions
implies that there are no issues with confounding between $\{\bolds{\beta
}_{t}\}$ and $\{\bu_{t}\}$; however, depending on our choice for $\{
\mathbf{M}_{t}\}$, there might be issues with confounding between $\bolds
{\eta}_{t-1}$ and the $(p+r)$-dimensional random vector $\bolds{\zeta}_{t}
\equiv(\bolds{\beta}_{t}^{\prime}, \bu_{t}^{\prime})^{\prime}$; $t =
2,\ldots,T$ [although the VAR(1) model assumes $\mathbf{u}_{t}$ is
independent of $\bolds{\eta}_{t-1}$]. Then, rewriting (\ref
{matrixprocess}), we get
%
\begin{equation}
\label{matrixprocess2} \mathbf{S}_{X,t}^{\prime}(\by_{t}-\bolds{
\xi}_{t}) = \mathbf {B}_{t}\bolds {\zeta}_{t} +
\mathbf{M}_{t}\bolds{\eta}_{t-1};\qquad t = 2,\ldots, T,
\end{equation}
where the $r\times(p+r)$ matrix $\mathbf{B}_{t}\equiv(\mathbf
{S}_{X,t}^{\prime}\mathbf{X}_{t}, \mathbf{I})$. The representation in
(\ref{matrixprocess2}) gives rise to what we call the MI propagator
matrix, which is defined in an analogous manner to the MI basis
functions. Using the spectral representation of $\mathbf{G}(\mathbf
{B}_{t},\mathbf{I}_{r}) = \bolds{\Phi}_{G,B,t}\bolds{\Lambda}_{G,B,t}\bolds
{\Phi
}_{G,B,t}^{\prime}$, we set the $r\times r$ real matrix $\mathbf
{M}_{t}$ equal to the first $r$ columns of $\bolds{\Phi}_{G,B,t}$ for each
$t$, which is denoted with $\mathbf{M}_{B,t}$.

Notice that there are no restrictions on $\{\mathbf{M}_{B,t}\}$ to
{mathematically guarantee} that $\mathbf{M}_{B,t}$ does not become
``explosive'' as $t$ increases. Thus, one should investigate whether or
not this is the case when using this model for ``long-lead''
forecasting. One should also be aware that we do not treat $\{\mathbf
{M}_{t}\}$ as an unknown parameter matrix to be estimated. Instead, we
chose a specific form for $\{\mathbf{M}_{t}\}$, namely, $\{\mathbf
{M}_{B,t}\}$, that avoids confounding between $\{\bolds{\eta}_{t}\}$ and
$\{\bolds{\zeta}_{t}\}$. As a result, the final form of $\{\mathbf
{M}_{B,t}\}$ might not be spatially interpretable. This issue is
addressed in Section~\ref{secparmod}, where constraints are added to
the parameter model so that $\operatorname{cov}(\bolds{\eta}_{t}) = \mathbf
{M}_{B,t}\mathbf{K}_{t-1}\mathbf{M}_{B,t}^{\prime} + \mathbf
{W}_{t}$ is
{spatially interpretable.} Nevertheless, it is a huge advantage in
spatio-temporal modeling to have a known propagator matrix, as a
prominent historical challenge with such models is addressing the curse
of dimensionality in estimating realistic propagators [\citet{cressie-wikle-book}, Chapter~7].

\subsection{Parameter models}\label{secparmod}


Methods for analyzing high-dimensional data (like the QWIs) seek to
remove ineffectual or redundant information [for a more in-depth
discussion see \citet{reviewmethods}]. In Sections~\ref{secmibf} and~\ref{secmipr} we
impose a reduced rank structure and a nonconfounding property and, as a
result, remove information on high frequencies and confounded random
effects, respectively. Thus, we specify $\{\mathbf{K}_{t}\}$ and $\{
\mathbf{W}_{t}\}$ in a manner that offsets these needed computational
compromises.

As an example, consider the case where we do not remove confounded
random effects. Let $\mathbf{P}_{X,t}\equiv\mathbf{X}_{t}(\mathbf
{X}_{t}^{\prime}\mathbf{X}_{t})^{-1}\mathbf{X}_{t}$ and the column
space of $\mathbf{P}_{X,t}$ be denoted as $\mathcal{C}(\mathbf
{P}_{X,t})$. Rewrite (\ref{processmodel}) and let $\mathbf{S}_{t} =
[\mathbf{H}_{X,t}, \mathbf{L}_{X,t}]$ and $\bolds{\eta}_{t}\equiv(\bolds
{\kappa}_{X,t}^{\prime},\bolds{\delta}_{X,t}^{\prime})^{\prime}$ so that
%
\begin{equation}
\label{decomp2} \by_{t} = \mathbf{X}_{t}\bolds{
\beta}_{t} + \mathbf{H}_{X,t}\bolds {\kappa }_{X,t} +
\mathbf{L}_{X,t}\bolds{\delta}_{X,t}+ \bolds{
\xi}_{t}; \qquad t = 2,\ldots,T.
\end{equation}
Here, the $N_{t}\times h$ matrix $\mathbf{H}_{X,t}\in\mathcal
{C}(\mathbf{P}_{X,t})^{\perp}$, the $N_{t}\times l$ matrix $\mathbf
{L}_{X,t}\in\mathcal{C}(\mathbf{P}_{X,t})$, $h$ and $l$ are
nonnegative integers, $\bolds{\kappa}_{X,t}$ is a $h$-dimensional
Gaussian random vector, and $\bolds{\delta}_{X,t}$ is a $l$-dimensional
Gaussian random vector; $t = 2,\ldots,T$. The decomposition in (\ref
{decomp2}) is the space--time analogue of the decomposition used for
discussion in \citet{Reich} and \citet{hughes}. The use of MI basis
functions is equivalent to setting $h$ equal to $r$, $\mathbf{H}_{X,t}
= \mathbf{S}_{X,t}$, and $\mathbf{L}_{X,t}$ equal to a $n_{t}\times l$
matrix of zeros for each $t$. As a result, the model based on MI basis
functions ignores the variability due to $\{\bolds{\delta}_{X,t}\}$
because it is confounded with $\{\bolds{\beta}_{t}\}$. In a similar manner,
one can argue that both the reduced rank structure of the MI basis
functions and the MI propagator matrix may also ignore other sources of
variability.

To address this concern, we consider specifying $\{\mathbf{K}_{t}\}$ as
positive semi-definite matrices that are ``close'' to target precision
matrices (denoted with $\mathbf{P}_{t}$ for $t = 1,\ldots,T$) that do
not ignore these sources of variability; critically, the use of a
target precision matrix allows us to reduce the parameter space in a
manner that respects the true variability of the process. Specifically,
let $\mathbf{K}_{t} = \sigma_{K}^{2}\mathbf{K}_{t}^{*}(\mathbf
{P}_{t})$, where $\sigma_{K}^{2}>0$ is unknown and
%
\begin{equation}
\label{Kstar} \mathbf{K}_{t}^{*}(\mathbf{P}_{t})=
\mathop{\operatorname{arg}\min}_{\mathbf{C}} \bigl\lbrace \bigl\Vert
\mathbf{P}_{t} - \mathbf {S}_{X,t}\mathbf{C}^{-1}
\mathbf{S}_{X,t}^{\prime}\bigr\Vert_{\mathrm
{F}}^{2} \bigr
\rbrace;\qquad t = 1,\ldots,T.
\end{equation}
Here, $\Vert\cdot\Vert_{\mathrm{F}}$ denotes the Frobenius norm. In (\ref
{Kstar}), we minimize the Frobenius norm across the space of positive
semi-definite matrices. A computable expression of $\mathbf
{K}_{t}^{*}(\mathbf{P}_{t})$ in (\ref{Kstar}) can be found in Appendix~\ref{appA}.

Processes with precision $\mathbf{P}_{t}$ do not ignore sources of
variability like $\bolds{\delta}_{X,t}$ in (\ref{decomp2}), since
$\mathbf
{P}_{t}$ has principal components in $\mathcal{C}(\mathbf{X}_{t})$ and
principal components associated with high frequencies. Hence, to
mitigate the effect of removing certain principal components when
defining $\mathbf{S}_{t}^{(\ell)}(\cdot)$, we specify the $r\times r$
matrix $\mathbf{K}_{t}$ to be as close as possible [in terms of the
Frobenius norm in (\ref{Kstar})] to something that has these principal
components, namely, the $n_{t}\times n_{t}$ matrix $\mathbf{P}_{t}$.
{That is, we rescale the total variability of our prior covariance to
account for variability ignored for reasons of computation and confounding.}

There are many choices for the ``target precision'' matrices $\{\mathbf
{P}_{t}\}$ in (\ref{Kstar}). For example, one might use a CAR model and
let $\mathbf{P}_{t} = \mathbf{Q}_{t}$, where recall $\mathbf{Q}_{t} =
\mathbf{I}_{N_{t}} - \mathbf{A}_{t}$; $t = 1,\ldots,T$. This allows one
to incorporate neighborhood information into the priors for $\{\mathbf
{K}_{t}\}$. In the case where the areal units are small and regularly
spaced, one might consider one of the many spatio-temporal covariance
functions that are available [e.g., see \citet{Gneitingcorr}, \citet
{HuangCressie2}, and \citet{steinSep}]. Alternatively, an empirical
Bayesian approach might be considered and an estimated precision (or
covariance) matrix might be used [e.g., see \citet{guttorpandpeterson}].

The spatial-only case provides additional motivation for the approach
in (\ref{Kstar}). That is, when $T = L = 1$ and $\mathbf{P}_{1} =
\mathbf{Q}_{1}$, the prior specification in (\ref{Kstar}) yields the MI
prior introduced in \citet{hughes}. This motivating special case is
formally stated and shown in Appendix~\ref{appA}.

With both $\{\mathbf{K}_{t}\}$ and $\{\mathbf{M}_{t}\}$ specified we
can solve for $\{\mathbf{W}_{t}\}$, that is, using the VAR(1) model
%
\begin{equation}
\label{Wstar} \mathbf{W}_{t} = \mathbf{K}_{t} -
\mathbf{M}_{B,t}\mathbf {K}_{t-1}\mathbf{M}_{B,t}^{\prime}
\equiv\sigma_{K}^{2}\mathbf {W}_{t}^{*};
\qquad t = 2,\ldots,T.
\end{equation}
In (\ref{Wstar}), the $r\times r$ matrix $\mathbf{W}_{t}^{*} =
\mathbf
{K}_{t}^{*} - \mathbf{M}_{B,t}\mathbf{K}_{t-1}^{*}\mathbf
{M}_{B,t}^{\prime}$; $t = 2,\ldots,T$. It is important to note that the
$r\times r$ matrices in the set $\{\mathbf{W}_{t}^{*}\}$ may not be
necessarily positive semi-definite. If $\mathbf{W}_{t}^{*}$ is not
positive semi-definite for some $t$, then we suggest using the best
positive approximation. This is similar to ``lifting'' adjustments
suggested by \citet{kang-cressie-shi-2010} in the spatio-temporal setting.

The prior distributions for the remaining parameters are specified so
that conjugacy can be used to obtain exact expressions for the full
conditionals within a Gibbs sampling algorithm. Specifically, we choose
a Gaussian distribution for $\{\bolds{\beta}_{t}\}$ and inverse gamma (IG)
for $\sigma_{K}^{2}$ and $\{\sigma_{\xi,t}^{2}\}$. In many cases the
statistical agency will provide values for $\{v_{t}^{(\ell)}(A)\}$ and,
thus, no model is required for $\{v_{t}^{(\ell)}(A)\}$ in this setting.
For our motivating QWI example, the LEHD program provides imputation
variances for QWIs (\url{http://download.vrdc.cornell.edu/qwipu.experimental/qwiv/beta1/}).
Imputation variances for QWIs are not available for each
county/quarter/industry/\break gender combination, which is the
multivariate spatio-temporal support of the data in Section~\ref{secmotiv}. Thus, we use an IG prior based on the imputation variances
that are available. See Appendix~\ref{appB} for details regarding the MCMC
algorithm, a complete summary of our statistical model, and a
discussion on alternative model specifications for related settings.

\section{Analysis of quarterly workforce indicators using the
MSTM}\label{analysis}

In this section we use the MSTM to analyze quarterly average monthly
income. In particular, our analysis has two primary goals. The first
goal is to demonstrate that the MSTM can reasonably reproduce latent
multivariate spatio-temporal fields for the QWI setting. To do this, we
perform an ``empirical study.'' Specifically, we perturb a subset of
the log quarterly average monthly income (log QWIs), introduced in
Section~\ref{secmotiv}, then we test whether or not we can recover the
log QWIs using the perturbed version. (Notice that the symmetrizing log
transformation is used so that the Gaussian assumptions from
Section~\ref{secMSTM} are met.) An empirical study such as this
differs from a traditional simulation study since the emphasis is on
illustrating that the MSTM can reproduce values similar to quarterly
average monthly income. Therefore, in Section~\ref{s51} we introduce our
empirical study design and in Section~\ref{s52} we provide the results of our
empirical study.

Our second goal in this section is to establish that the MSTM can be
efficiently used to jointly model high-dimensional areal data (see
Section~\ref{secmotiv} for a discussion). The methodological
development in Sections~\ref{secMSTM} and~\ref{secspecs} are
motivated by striking a balance between modeling realistic multivariate
spatio-temporal dependencies and allowing for the possibility of
extremely high-dimensional data sets. As such, in Section~\ref{secmassive} we
jointly analyze all 7{,}530{,}037 quarterly average monthly income
estimates provided by the LEHD program.

For Sections~\ref{s51} through \ref{secmassive}, the Gibbs sampler, provided in
Appendix~\ref{appB}, was run for 10{,}000 iterations with a burn-in of 1000 iterations.
Convergence of the Markov chain Monte Carlo algorithm was assessed
visually using trace plots of the sample chains, with no lack of
convergence detected. {Additionally, the batch means estimate of the
Monte Carlo error (with batch size 50) [e.g., see \citet{batch1,batch2}] and the Gelman--Rubin diagnostic (computed using
three chains) [e.g., see \citet{gr1}] did not suggest lack of convergence.}

\subsection{Empirical study design}\label{s51}
Abowd et~al. (\citeyear{abowd}) provide a study to
assess the quality of the QWIs. Thus, for consistency within the
literature we adopt a study design similar to the one used in
Section~5.7.2 of \citet{abowd}.
Specifically, we restrict the data to $t =
4,\ldots,55$ (quarters between 1991 and 2003), $\ell= 1,2$ (which
represents women and men in the education industry, respectively), and
the prediction locations equal the counties in Minnesota that have
available QWIs (i.e., $D_{\mathrm{P},t}^{(\ell)} \equiv D_{\mathrm
{MN},t}^{(\ell)}$). The scope of this empirical study is smaller than
the entire data set introduced in Section~\ref{secmotiv}, since in
this section we are primarily interested in showing that the MSTM can
recover latent multivariate spatio-temporal fields similar to the
quarter average monthly income. See Section~\ref{secmassive} for a
demonstration of using the MSTM to efficiently jointly model the entire
7{,}530{,}037 QWIs.

The perturbed version of the log quarterly average monthly income is
explicitly written as
%
\begin{equation}
\label{perturb}\hspace*{8pt} R_{t}^{(\ell)}(A) = Z_{t}^{(\ell)}(A)
+ \epsilon_{t}^{(\ell
)}(A);\qquad t = 4,\ldots, 55, \ell= 1,2, A
\in D_{\mathrm{MN},t}^{(\ell)},
\end{equation}
where $D_{\mathrm{MN},t}^{(\ell)}$ is the set of counties in Minnesota
(MN) that have available quarterly average monthly income estimates, $\{
R_{t}^{(\ell)}(A)\}$ represents the perturbed version of the log
quarterly average monthly income [log QWIs; denoted by $\{Z_{t}^{(\ell
)}(\cdot)\}$], and the set $\{\epsilon_{t}^{(\ell)}(A): t = 4,\ldots,
55, \ell= 1,2, A \in D_{\mathrm{O},t}^{(\ell)}\}$ consists of i.i.d.
normal random variables with mean zero and variance $\sigma_{\epsilon
}^{2}$. In practice, the quarterly average monthly income estimates are
publicly available and are, hence, observed. Nevertheless, for the
purposes of this empirical study we will act as if the QWIs are an
unobserved multivariate spatio-temporal field to be estimated, and
treat $\{R_{t}^{(\ell)}\}$ as the data process and $\{Z_{t}^{(\ell
)}(\cdot)\}$ as the latent process.

We randomly select 65$\%$ of the areal units in $D_{\mathrm
{MN},t}^{(\ell)}$ to be ``observed,'' which we denote with the set
$D_{\mathrm{MN, O}, t}^{(\ell)}$. {Thus, for this example,
$D_{\mathrm
{P},t}$ (given by $D_{\mathrm{MN},t}^{(\ell)}$) and $D_{\mathrm{O},t}$
(given by $D_{\mathrm{MN, O}, t}^{(\ell)}$) are not the same.} Recall
from Section~\ref{secmotiv} that this choice reflects the amount of
observed data present in the entire QWI data set, where 65$\%$ of the
QWIs are observed. However, it is important to note that the ``missing
QWI'' structure of the data set in Section~\ref{secmotiv} is different
from what we use in this empirical study, since we do not incorporate
missing QWIs patterns that occur due to a state's failure to sign a
MOU. Recall that if a state does not sign a MOU for a particular year,
then the entire state is missing for that year. However, our choice to
randomly select 65$\%$ of the areal units within $D_{\mathrm
{MN},t}^{(\ell)}$ to be ``observed'' is sufficient for our purposes.

The value for the perturbation variance $\sigma_{\epsilon}^{2}$ is
chosen relative to the variability of the log quarterly average monthly
income. The variance of the log quarterly average monthly income,
within\vspace*{1pt} our study region, is given by $\operatorname{var}\{Z_{t}^{(\ell)}(A)\} =
0.24$. Thus, we specify the perturbations $\{\epsilon_{t}^{(\ell)}(A):
A \in D_{\mathrm{O},t}^{(\ell)}\}$ to have variance $\sigma
_{\epsilon
}^{2}\equiv0.24$. This yields a signal-to-noise ratio of 1, which can
be interpreted as a small signal-to-noise ratio. We argue that this
choice is conservative, since small signal-to-noise ratios
traditionally make prediction of a latent process difficult [\citet
{cressaldworth}].

\begin{figure}

\includegraphics{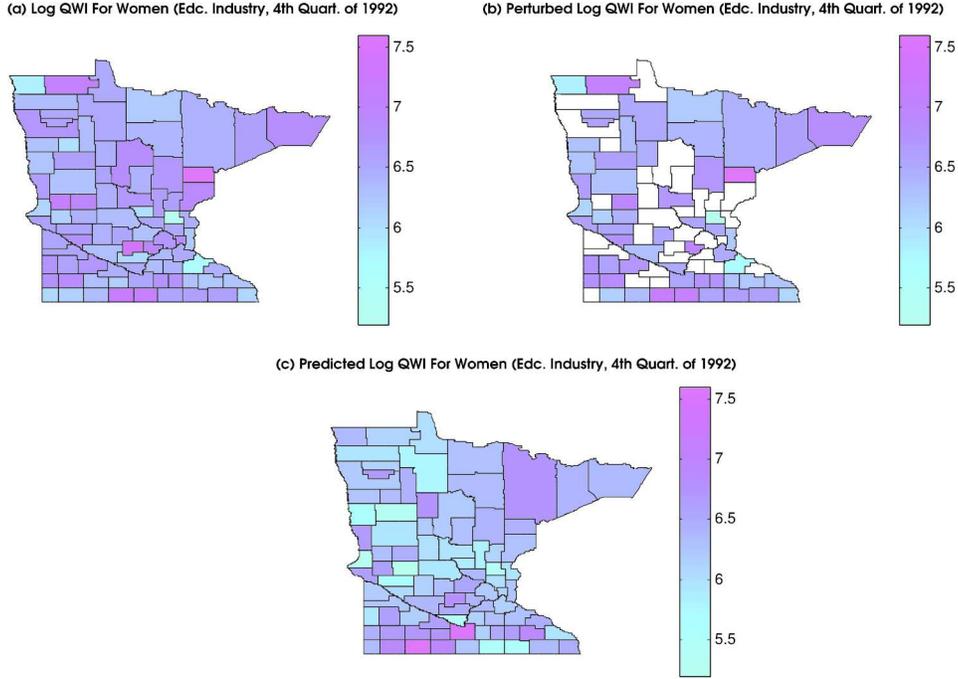}

\caption{\textup{(a)} Map of the log quarterly average
monthly income for women in the education industry [i.e., $\{
Z_{8}^{(1)}(A)\}$]. These values correspond to log quarterly average
monthly income for women in the education industry, for counties in
Minnesota, and for the 4th quarter in 1992. For comparison, a~map of
the perturbed log quarterly average monthly income for women in the
education industry [i.e., $\{W_{8}^{(1)}(A)\}$] is given in \textup{(b)}. The
white areas indicate missing regions. In \textup{(c)} we provide the predictions
$\{\widehat{Z}_{8}^{(1)}(A)\}$ that are computed using MSTM and the
perturbed data $\{R_{t}^{(\ell)}(A)\}$ from equation (\protect\ref{perturb}).}\label{fig2}
\end{figure}

We end this section with an example of analyzing a single realization
of $\{R_{t}^{(\ell)}(A): t = 4,\ldots, 55, \ell= 1,2, A \in
D_{\mathrm
{MN,O},t}^{(\ell)}\}$. Consider the selected maps of the log quarterly
average monthly income and the perturbed log average monthly income in
Figure~\ref{fig2}(a) and~(b), respectively. Figure~\ref{fig2} visually depicts the
difficulty of predicting a latent random field, {as} {the number of
``missing'' QWIs is rather large and the signal-to-noise ratio is
visibly small.}

\begin{figure}

\includegraphics{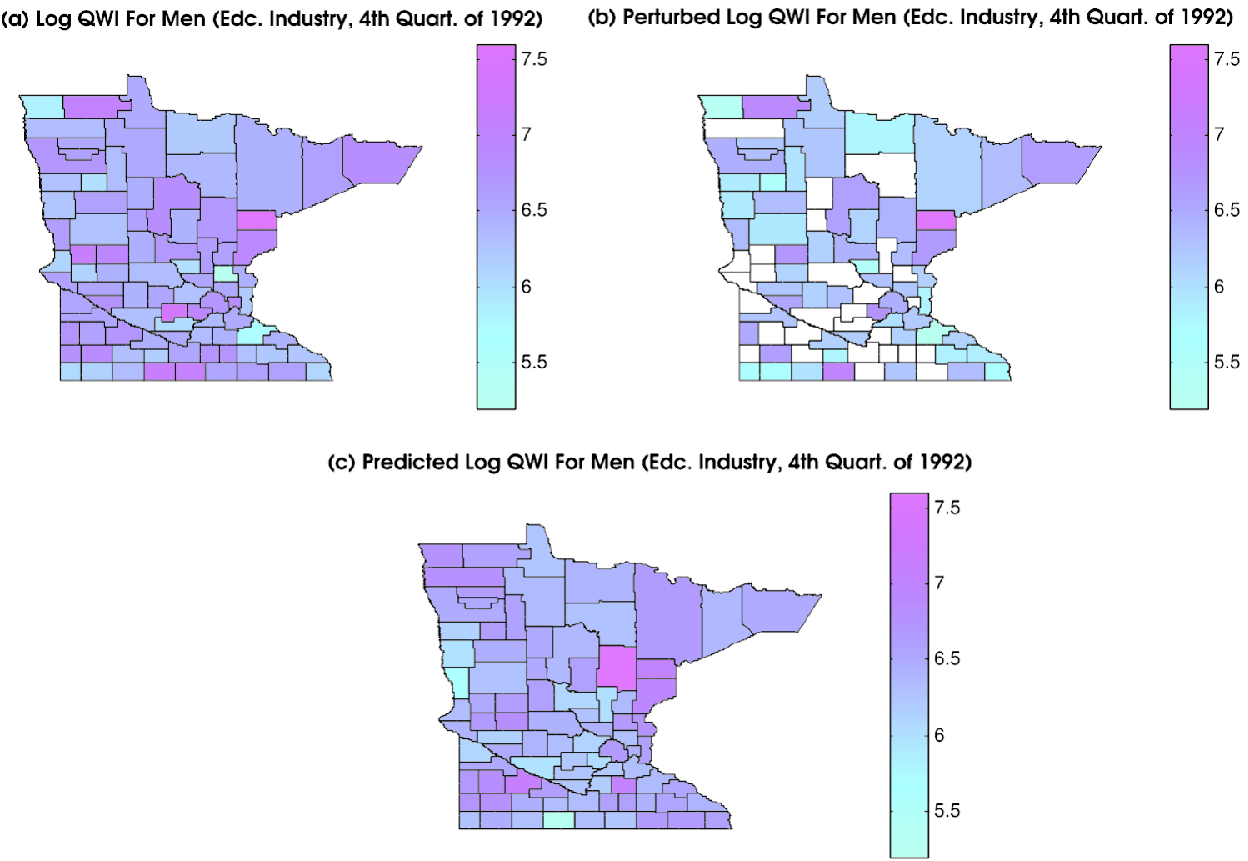}

\caption{\textup{(a)} Map of the log quarterly average
monthly income for men in the education industry [i.e., $\{
Z_{8}^{(2)}(A)\}$]. These values correspond to log quarterly average
monthly income for men in the education industry, for counties in
Minnesota, and for the 4th quarter in 1992. For comparison, a map of
the perturbed log quarterly average monthly income for men\vspace*{1pt} in the
education industry [i.e., $\{W_{8}^{(2)}(A)\}$] is given in \textup{(b)}.\vspace*{1pt} The
white areas indicate missing regions. In \textup{(c)} we provide the predictions
$\{\widehat{Z}_{8}^{(2)}(A)\}$ that are computed using MSTM and the
perturbed data $\{R_{t}^{(\ell)}(A)\}$ from equation (\protect\ref{perturb}).}
\label{fig3}\vspace*{-6pt}
\end{figure}

To use the MSTM to predict $\{Z_{t}^{(\ell)}\}$ from $\{R_{t}^{(\ell
)}\}
$, we need to specify the target precision matrix, the covariates, and
the number of MI basis functions. Set the target precision matrix equal
to $\{\mathbf{Q}_{t}\}$ as previously described below {(\ref{Kstar})}.
Let $\mathbf{x}_{t}^{(\ell)}(A)\equiv1$, where $g = 1,2$ indexes men
and women, respectively. Also, for illustration let $r=30$, which is
roughly 50$\%$ of the available MI basis functions at each time point
$t$. In a sensitivity study (not shown), we see that the MSTM is
relatively robust to changes to larger values of $r$. In general, for
the purposes of prediction, large values of $r$ are preferable;
however, a carefully selected reduced rank set of basis functions can
produce as good or better predictions than those based on the full set
of basis functions [\citeauthor{bradley2011} (\citeyear{bradley2011,bradley2014comp,bradleyTEST})].
Using the MSTM with these specifications, we predict $Z_{t}^{(\ell)}$
using the perturbed values $R_{t}^{(\ell)}$. In Figure~\ref{fig2}(c) we present
$\{\widehat{Z}_{8}^{(1)}(A): A \in D_{\mathrm{O},8}^{(1)}\}$. In
general, we let $\widehat{Z}_{t}^{(\ell)}$ denote the MSTM predictions
based on $\{R_{t}^{(\ell)}(A): t = 4,\ldots,55, \ell= 1,2, A \in
D_{\mathrm{MN,O},t}^{(\ell)}\}$. Similar conclusions are drawn from
Figure~\ref{fig3}, which provides results for men.

\begin{figure}

\includegraphics{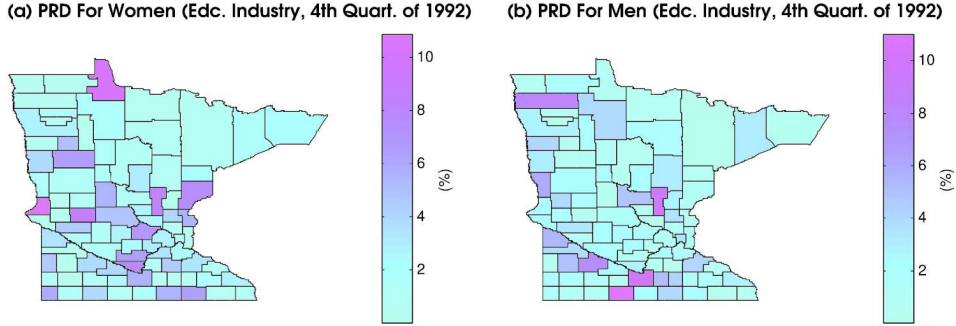}

\caption{In \textup{(a)} and \textup{(b)}, the percent relative
difference (PRD) in (\protect\ref{mprdone}) of the predicted log quarterly
monthly average income of women and men within the education industry
during the fourth quarter of 1992.}
\label{fig4}
\end{figure}

The performance of our predictions are further corroborated by the
results presented in Figure~\ref{fig4}(a) and (b), where we map the percent
relative difference (PRD) between the predicted log quarterly average
monthly income and the actual log quarterly average monthly income.
That is, the values plotted in {Figure~\ref{fig4}(a) and (b)} are given by
%
\begin{equation}
\label{mprdone} \operatorname{abs} \biggl\lbrace\frac{\widehat{Z}_{8}^{(\ell
)}(A)-Z_{8}^{(\ell
)}(A)}{Z_{8}^{(\ell)}(A)} \biggr\rbrace
\times100\%;\qquad \ell= 1,2, A \in D_{\mathrm{O},8}^{(\ell)}.
\end{equation}
Additionally, the median PRD across all variables, regions, and time
points is 4.87$\%$. Hence, for this example the difference between the
predicted and actual log quarterly average monthly income is small
relative to the scale of the log quarterly average monthly income.
{Thus, we appear to be efficiently reproducing the unobserved latent
field (as measured by PRD) using the MSTM}.

\subsection{Empirical study of multiple replicates}\label{s52}
There have been no
statistical methods used to obtain QWI estimates and measures of
precision at missing regions. Thus, in this section we evaluate the
performance of $\{\widehat{Z}_{t}^{(\ell)}\}$ at both observed and
missing regions over multiple replicates.

\begin{figure}

\includegraphics{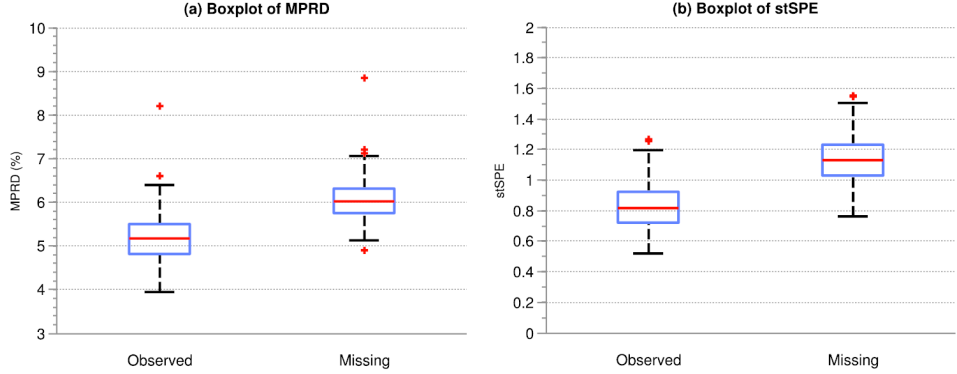}

\caption{\textup{(a)} Boxplots of the MPRD in (\protect\ref
{mprdemp}), using the 50 replicates of the spatial field $\{
R_{t}^{(\ell)}(A): t = 4,\ldots,55, \ell= 1,2, A \in D_{\mathrm
{MO,O},t}^{(\ell)}\}$, for observed and missing respectively. \textup{(b)}
Boxplots of the {stSPE} in (\protect\ref{mspeemp}), using the 50 replicates of
the spatial field $\{R_{t}^{(\ell)}(A): t = 4,\ldots,55, \ell= 1,2, A
\in D_{\mathrm{MO,O},t}^{(\ell)}\}$, for observed and missing
respectively. In both \textup{(a)} and \textup{(b)} we do not plot outliers for the
purposes of visualization.}\label{fig5}
\end{figure}

The MSTM from Section~\ref{secMSTM} is currently the only stochastic
modeling approach available to jointly model high-dimensional
multivariate spatio-temporal areal data. Since there are no viable
alternative methods available, we first assess the quality of the
predictions relative to the scale of the data [e.g., see equation (\ref
{mprdone})]. Specifically, consider the median percent relative
difference (MPRD) given by
%
\begin{eqnarray}
\mathrm{MPRD}  &\equiv &  \operatorname{median} \biggl\lbrace
\operatorname{abs} \biggl[\frac
{\widehat{Z}_{t}^{(\ell)}(A)-Z_{t}^{(\ell)}(A)}{Z_{t}^{(\ell
)}(A)} \biggr] \times100:
\nonumber
\\[-8pt]
\label{mprdemp}
\\[-8pt]
\nonumber
 && {}t = 4,\ldots,55, \ell=
1,2, A \in D_{\mathrm
{O},t}^{(\ell)} \biggr\rbrace.
\end{eqnarray}
If MPRD in (\ref{mprdemp}) is ``close'' to zero for a given replicate
of the field $\{R_{t}^{(\ell)}(A): t = 4,\ldots, 55, \ell= 1,2, A
\in
D_{\mathrm{MN,O},t}^{(\ell)}\}$, then the predictions are considered
close (relative to the scale of the data) to the log quarterly average
monthly income. In Figure~\ref{fig5}(a) we provide boxplots [over 50 independent
replicates of $\{R_{t}^{(\ell)}\}$] of MPRD evaluated at observed and
missing regions, respectively. Here, we see that the MPRD is larger at
missing regions as expected. However, the values of the MPRD are
consistently small for both observed and missing regions: the medians
are given by 5.17$\%$ and 6.02$\%$ for observed and missing regions,
respectively; and the interquartile ranges are given by 0.6915 and
0.5470 for observed and missing regions, respectively. Thus, the MPRD
shows that we are obtaining predictions that are close (relative to the
scale of the log QWIs) to the log quarterly average monthly income.

Another metric that one might use to validate our conclusions from
Figure~\ref{fig5}(a) is the standardized squared prediction error {(stSPE)}
%
\begin{eqnarray}
 \mathrm{{stSPE}}& =&  \operatorname{average} \bigl\lbrace\bigl(
\widehat {Z}_{t}^{(\ell
)}(A)-Z_{t}^{(\ell)}(A)
\bigr)^{2}:
\nonumber
\\[-8pt]
\label{mspeemp}
\\[-8pt]
\nonumber
 &&{}t = 4,\ldots,55, \ell= 1,2, A \in D_{\mathrm{O},t}^{(\ell)}
\bigr\rbrace/\sigma_{\epsilon}^{2}.
\end{eqnarray}
If {stSPE} in (\ref{mspeemp}) is ``close'' to zero for a given
replicate of the field $\{R_{t}^{(\ell)}(A): t = 4,\ldots, 55, \ell=
1,2, A \in D_{\mathrm{MN,O},t}^{(\ell)}\}$, then the predictions are
considered close to the log quarterly average monthly income. Also
notice that the {stSPE} in (\ref{mspeemp}) is normalized by $\sigma
_{\epsilon}^{2}$; consequently, we can compare the squared error of our
predictions relative to the perturbation variances. This is especially
noteworthy for predictions at missing regions, which have no signal in
the original perturbed data set.

In Figure~\ref{fig5}(b) we provide boxplots [over 50 independent replicates of
$\{R_{t}^{(\ell)}\}$] {stSPE} evaluated at observed and missing
regions, respectively. Here, we see that the MSPE is larger at missing
regions as expected. However, the values of the {stSPE} at observed
(missing) regions are consistently smaller (close) than 1: the medians
are given by 0.8154 and 1.1293 for observed and missing regions,
respectively; and the interquartile ranges are given by 0.1994 and
0.1990 for observed and missing regions, respectively. Thus, the
{stSPE} shows that the error in our predictions at observed (missing)
regions are smaller than (similar to) the perturbation error (i.e.,
$\sigma_{\epsilon}^{2}$).

{Notice that the stSPE is roughly 0.1293 above 1 at missing locations
and 0.1846 below 1 at observed locations; thus{,} the relative
differences from 1 are similar in the two situations. This may be
problematic if {there are} more missing values than observed. However,
note that this is not the case for the LEHD data set, which has roughly
65$\%$ of the prediction locations observed.}

\subsection{Predicting quarterly average monthly income}\label{secmassive} 
\begin{figure}

\includegraphics{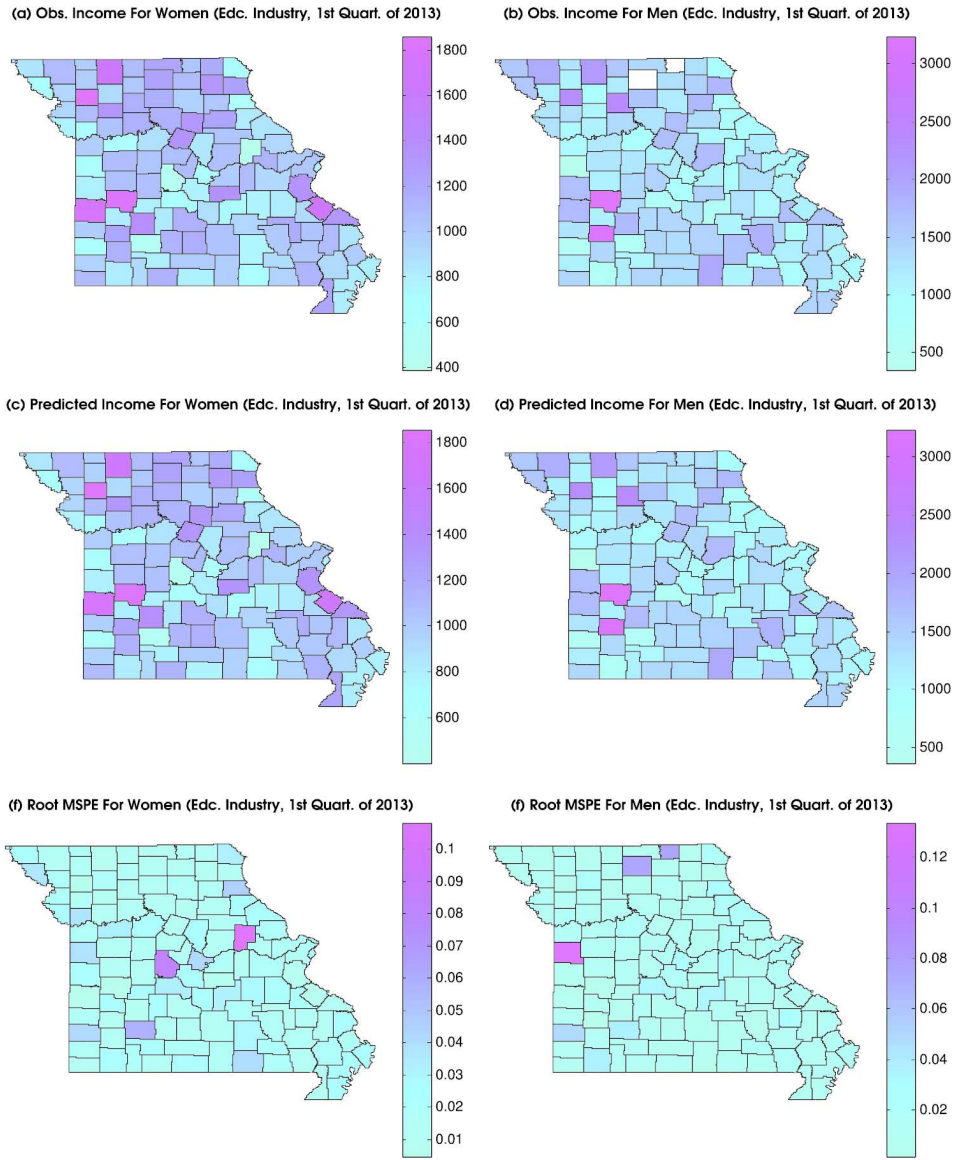}

\caption{\textup{(a)} and \textup{(b)} present the QWI for quarterly
average monthly income ({US dollars}) for the state of Missouri, for
each gender, for the education industry, and for the first quarter of
2013. LEHD does not provide estimates at every county in the US at
every quarter; these counties are shaded white. \textup{(c)}--\textup{(f)} present the
corresponding maps (for the state of Missouri, for each gender, for the
education industry, and for quarter 92) of predicted monthly income
({US dollars}) and their respective posterior square root MSPE (on the
log-scale). Notice that the color scales are different for women and
men {and that the root MSPEs are computed on the log-scale}. White
areas indicate missing regions.}\label{fig6}
\end{figure}

We demonstrate the use of MSTM using a high-dimensional
multivariate spatio-temporal data set
made up of quarterly average monthly income obtained from the LEHD
program. In particular, we consider all 7{,}530{,}037 observations
introduced in Section~\ref{secmotiv}. These values are available over
the entire US, which we jointly analyze using the MSTM. We present a
subset of this data set in Figure~\ref{fig6}(a) and~(b). We see that the
quarterly average monthly income is relatively constant across each
county of the state of Missouri and that men tend to have higher
quarterly average monthly income than women. This pattern is consistent
across the different spatial locations, industries, and time points.

%

The primary goals of our analysis in this section is to estimate the
quarterly average monthly income, {investigate potential gender
inequality in the US}, and determine whether or not it is
computationally feasible to use the MSTM for a data set of this size.
Preliminary analyses indicate that the log quarterly average monthly
income is roughly Gaussian. Since we assume that the underlying data is
Gaussian, we treat the log of the average income as $\{Z_{t}^{(\ell
)}(\cdot)\}$ in (\ref{datamodel}).

For illustration, we make the following specifications. Set the target
precision matrix equal to $\{\mathbf{Q}_{t}\}$ as previously described
below (\ref{Kstar}). Let $\mathbf{x}_{t}^{(\ell)}(A)\equiv(1,I(\ell=
1),\ldots,I(\ell= 39), I(g = 1)\times I(\ell= 1),\ldots,I(g =
1)\times I(\ell= 39))^{\prime}$, where $g = 1,2$ indexes men and
women, respectively, and recall $I(\cdot)$ is the indicator function.
Also, following the MSTM specifications from our empirical study, we
let $r=30$, which is roughly 50$\%$ of the available MI basis functions
at each time point $t$. Using the MSTM with these specifications, we
predict $L\times T = 40\times92 = 3680$ different spatial fields. The
CPU time required to compute these predictions is approximately 1.2
days,\vadjust{\goodbreak} with all of our computations performed in Matlab (Version~8.0) on
a dual 10 core 2.8 GHz Intel Xeon E5-2680 v2 processor, with 256 GB of
RAM. Of course, additional efforts in efficient programming may result
in faster computing; however, these results indicate that it is
computationally practical to use the MSTM to analyze massive data.

Although we modeled the entire US simultaneously, for illustration, we
present maps of predicted monthly income for the state of Missouri, for
each gender, for the education industry, and for the 92-nd quarter
[Figure~\ref{fig6}(c) and~(d)]. The prediction maps are essentially constant
over the state of Missouri, where women tend to have a predicted
monthly income of slightly less than 1200 dollars and men consistently
have a predicted monthly income of about 1800 dollars. As observed in
Figure~\ref{fig6}(a) and~(b), there is a clear pattern where men have higher
predicted monthly income than women. These predictions appear
reasonable since the maps of the root MSPE {(on the log scale)}, in
Figure~\ref{fig6}(e) and~(f), indicate we are obtaining precise predictions
{on the log-scale}. Additionally, upon comparison of Figure~\ref{fig6}(a) and~(b) to Figure~\ref{fig6}(c) and~(d), we see that the predictions reflect the
same general pattern in the data. These results are similar across the
different states, industries, and time points.

To further corroborate the patterns in the MSTM predictions, we fit a
separate univariate spatial model from \citet{hughes}. Specifically, we
fit the univariate spatial model from \citet{hughes} to the data in
Figure~\ref{fig6}(a) and~(b) with $r=62$ basis functions ($100\%$ of the
available basis functions) and obtain the prediction maps (not shown).
Notably, the predictions are also fairly constant around 1200 and
1800 dollars. Moreover, the MSPE of the \citet{hughes} predictions
(summed over all US counties) is 4.09 times larger than the MSPE of the
predictions from the MSTM summed over all US counties. This may be due,
in part, to the fact that the model in \citet{hughes} does not
incorporate multivariate and serial (temporal) dependencies.

\begin{figure}

\includegraphics{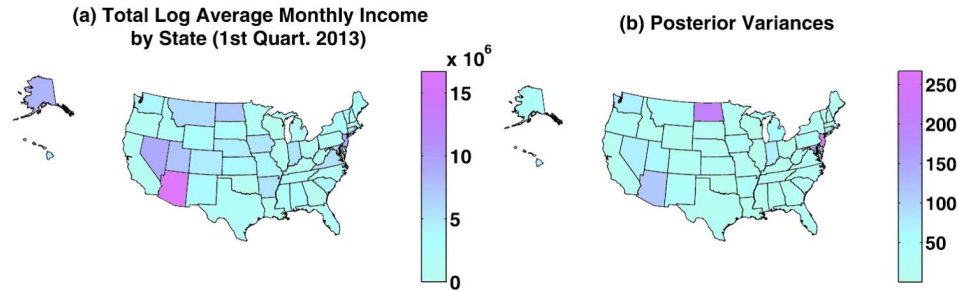}

\caption{Plots of contrasts of $\bolds{\mu}_{t}^{(\ell)}(A)$ (referred to as the log average monthly income). In \textup{(a)} and~\textup{(b)},
we plot\vspace*{1pt} the posterior mean and variance of $\sum\bolds{\mu
}_{92}^{(m_{k})}(A)-\sum\bolds{\mu}_{92}^{(w_{k})}(A)$ by state, where
the sum aggregates counties to states and is also indexed over the industries.}\label{fig7}
\vspace*{-3pt}
\end{figure}

The large difference in average monthly income between men and women
can be { further investigated} by comparing the means [i.e., $\bolds{\mu
}_{t}^{(\ell)}(\cdot)$] for men and women, respectively. [Recall from
Section~\ref{secmibf} that we can perform inference on $\bolds{\mu}_{t}^{(\ell
)}(\cdot)$ because we impose a nonconfounding property between $\bolds
{\mu
}_{t}^{(\ell)}(\cdot)$ and $\mathbf{S}_{t}^{(\ell)}(\cdot)^{\prime
}\bolds
{\eta}_{t}$.] Now, let $m_{1},\ldots,m_{20}$ indicate industry 1
through 20 for men, and $w_{1},\ldots,w_{20}$ for women. Then, for a
given $A$ consider the contrast given by $\sum_{k = 1}^{20}\mu
_{92}^{(m_{k})}(A)-\sum_{k = 1}^{20}\mu_{92}^{(w_{k})}(A)$, which is
interpreted as an average difference between the income of men and
women over the 20 industries. Hence, this contrast is a global (across
industries) measure of income gender differences at the most current
time point (notice $t = 92$). A positive (negative) value indicates
that men (women) tend to have larger incomes. In Figure~\ref{fig7}(a) and~(b)
we plot the posterior mean and variance of this contrast by state.
Here, we see that for the {first} quarter of 2013, gender inequality is
similar across each state (with men consistently { having} larger
quarterly incomes), with the largest disparity occurring in Arizona.

\begin{figure}

\includegraphics{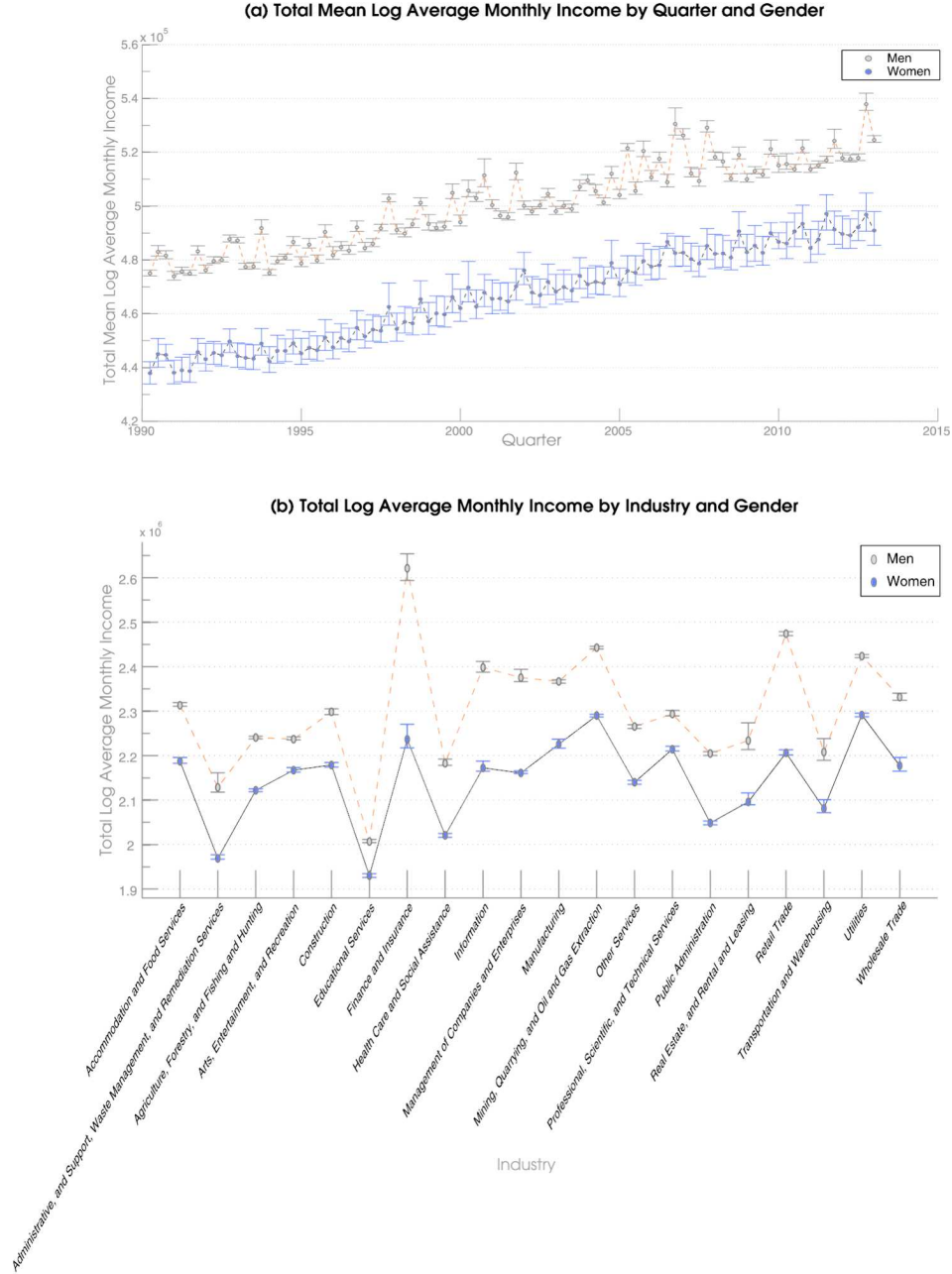}
\vspace*{-3pt}
\caption{Plots of contrasts of $\bolds{\mu}_{t}^{(\ell
)}(A)$ (referred to as the log average monthly income). In \textup{(a)}, we plot
the posterior mean of $\sum_{k,A} \bolds{\mu}_{t}^{(m_{k})}(A)$ and
$\sum_{k,A} \bolds{\mu}_{t}^{(w_{k})}(A)$ by quarter. In \textup{(b)}, we plot the
posterior mean of $\sum_{t,A} \bolds{\mu}_{t}^{(m_{k})}(A)$ and $\sum_{t,A} \bolds{\mu}_{t}^{(w_{k})}(A)$ by industry. In both \textup{(a)} and \textup{(b)} a
$95\%$ credible interval is given by horizontal line segments, and for
comparison a line is superimposed connecting the intervals associated
with males and females, respectively.}\label{fig8}
\end{figure}

Figure~\ref{fig7}(a) and~(b) give a sense of the spatial patterns of the
between-gender income differences for the {first} quarter of 2013. We
can also investigate the temporal and between-industry patterns in a
similar manner. In particular, in Figure~\ref{fig8}(a) we plot $\sum_{k,A} \bolds
{\mu}_{t}^{(m_{k})}(A)$ and $\sum_{k,A} \bolds{\mu}_{t}^{(w_{k})}(A)$ by
quarter (i.e., $t$). Here, we see that the differences between the
genders appears to be constant from 1990 to 2013. Likewise, in
Figure~\ref{fig8}(b) we identify between industry differences by plotting the
posterior mean of $\sum_{t,A} \bolds{\mu}_{t}^{(m_{k})}(A)$ and $\sum_{t,A} \bolds{\mu}_{t}^{(w_{k})}(A)$ by industry (i.e., $k$). Here, we
observe that gender inequality appears present in each industry, with
men consistently having larger mean average monthly income. That is,
the posterior mean of $\sum_{t,A} \bolds{\mu}_{t}^{(m_{k})}(A)$ and the
values within 95$\%$ (pointwise) credible intervals are larger than
that for women. Furthermore, we see that the largest difference between
log average monthly income occurs in the finance and insurance
{industries}, which also appear to be the most lucrative {industries}
for men.

It should be noted that, despite the inherent computational issues,
having an abundance of data has distinct advantages. For example,
notice in Figure~\ref{fig6}(b) that LEHD does not release data at two counties
of Missouri for men in the education industry during quarter 92.
Although these values are missing for this variable and time point,
LEHD releases QWIs at these two counties (for men in the education
industry) for 43 different quarters. Hence, with the observed values
from 43 different spatial fields, we reduce the variability of
predictions at the two missing counties during the 92nd quarter
[compare Figure~\ref{fig6}(b) to (f)]. This is particularly useful for the
setting when a states does not sign a MOU and, hence, LEHD does not
provide estimates\vadjust{\goodbreak} here.

\section{Discussion}\label{secdisc} We have introduced fully Bayesian
methodology to analyze areal data sets with multivariate
spatio-temporal dependencies. In particular, we introduce the
multivariate spatio-temporal mixed effects model (MSTM). {To date,
little} has been proposed to model areal data that exhibit multivariate
spatio-temporal dependencies. Furthermore, the available alternatives
[see \citet{carlinmst} and \citet{daniels}] do not allow for certain
complexities in cross-covariances and fail to accommodate
high-dimensional data sets. Hence, the MSTM provides an important
addition to the multivariate spatio-temporal literature.

The MSTM was motivated by the Longitudinal Employer-Household Dynamics
(LEHD) program's quarterly workforce indicators (QWI) [\citet{abowd}].
In particular, the QWIs are extremely high-dimensional and exhibit
complex multivariate spatio-temporal dependencies. Thus,\vadjust{\goodbreak} extensive
methodological contributions, leading to the MSTM, were necessary in
order to realistically, jointly model the QWIs' complex multivariate
spatio-temporal dependence structure and to allow for the possibility
of remarkably high-dimensional areal data.

We conducted an extensive empirical study to demonstrate that the MSTM
works extremely well for predicting the QWI, quarterly average monthly
income. Specifically, we perturb the log quarterly average monthly
income, then predictions of the log quarterly average monthly income
are made using the perturbed values and comparisons are made between
the predicted and the actual log quarterly average monthly income. The
results illustrate that we are consistently recovering the unobserved
latent field using the MSTM at both observed and missing regions. This
is particularly noteworthy, since there are no other methods that have
been used to estimate QWIs at missing regions. In fact, because we
borrow strength over different variables, space, and time, we can also
predict values for entire states when the values are missing for
reasons of an unsigned MOU.

The exceptional effectiveness of our approach is further illustrated
through a joint analysis of all the available quarterly average monthly
income estimates. This data set, comprised of 7,530,037 observations,
is used to predict 3680 different spatial fields consisting of all the
counties in the US. The recorded CPU time for this example was 1.2
days, which clearly indicates that it is practical to use the MSTM in
high-dimensional data contexts.

{In this article, we have found that incorporating different variables,
space, and time into an analysis is beneficial for two reasons. First,
one can leverage information from nearby (in space and time)
observations and related variables to improve predictions and, second,
there are inferential questions that are unique to multivariate
spatio-temporal processes. For example, in Section~\ref{secmassive} it was of
interest to determine where, when, and what industry had the largest
disparity between the average quarterly income of men and women. Here,
we found that these differences have been relatively constant over the
last two decades, { are} currently fairly constant over each state, and
{ are} the highest within the finance and insurance {industries}.}

Although our emphasis was on analyzing QWIs, our modeling framework
allows the MSTM to be applied to a wide array of data sets. For
example, the MSTM employs a reduced rank approach to allow for massive
multivariate spatio-temporal data sets. Additionally, the MSTM allows
for nonstationary and nonseparable multivariate spatio-temporal
dependencies. This is achieved, in part, through a novel propagator
matrix for a first-order vector autoregressive [VAR(1)] model, which we
call the MI propagator matrix. This propagator matrix is an extension
of the MI basis function [\citeauthor{griffith2000} (\citeyear{griffith2000,griffith2002,griffith2004}), \citet{griffith2007,hughes,aaronp}] from the spatial-only
setting to the multivariate spatio-temporal setting. We motivate both
the MI basis function and the MI propagator matrix as an approximation
to a target precision matrix, that allows for both computationally
efficient statistical inference and nonconfounding regression parameters.

Our model specification also allows for knowledge of the underlying
spatial process to be incorporated into the MSTM. Specifically, we
propose an extension of the MI prior to the spatio-temporal case. This
extension forces the covariance matrix of the random effect to be close
(in Frobenius norm) to a ``target precision'' matrix, which can be
chosen based on knowledge of the underlying spatial process.
Importantly, this contribution has broader implications, in terms of
reducing a parameter space, for defining informative parameter models
for high-dimensional spatio-temporal processes.

There are many opportunities for future research. For example, there
are many QWIs available that are recorded as counts, which do not
satisfy the Gaussian assumption even after a transformation. Thus, the
MSTM could be extended to the Poisson data setting. The parameter model
introduced in Section~\ref{secparmod} is also of independent interest.
In our applications, we let $\{\mathbf{Q}_{t}\}$ be the target
precision. However, one could conceive of various different ``target
precisions'' built from deterministic models (e.g., for atmospheric
variables). Another avenue {for} future research is to extend the MI
propagator matrix, beyond the VAR(1) specification. In fact, this
strategy could be easily used for many subject matter domains for other
time series models.


\begin{appendix}\label{app}
\section{Technical results}\label{appA}

\begin{prop}\label{prop1}
Let $\bolds{\Phi}_{k}$ be a generic $n \times r$
real matrix such that $\bolds{\Phi}_{k}^{\prime}\bolds{\Phi}_{k}=\mathbf
{I}_{r}$, $\mathbf{C}$ be a generic $r \times r$ positive definite
matrix, $\mathbf{P}_{k}$ be a generic $n\times n$ positive definite
matrix, and let $k = 1,\ldots,K$. Then, the value of $\mathbf{C}$ that
minimizes $\sum_{k=1}^{K}\Vert\mathbf{P}_{k} - \bolds{\Phi}_{k}\mathbf
{C}^{-1}\bolds{\Phi}_{k}^{\prime}\Vert_{F}^{2}$ within the space of positive
semi-definite covariances is given by
\renewcommand{\theequation}{A.1}
\begin{equation}
\label{minfrobs} \mathbf{C}^{*} = \Biggl\lbrace\mathcal{A}^{+}
\Biggl(\frac{1}{K}\sum_{k =
1}^{K}
\bolds{\Phi}_{k}^{\prime}\mathbf{P}_{k}\bolds{
\Phi}_{k} \Biggr) \Biggr\rbrace^{-1},
\end{equation}
where $\mathcal{A}^{+}$($\mathbf{R}$) is the best positive approximate
[\citet{Higham}] of a real square matrix $\mathbf{R}$. Similarly, the
value of $\mathbf{C}$ that minimizes $\sum_{k=1}^{K}\Vert\mathbf{P}_{k} -
\bolds{\Phi}_{k}\mathbf{C}\bolds{\Phi}_{k}^{\prime}\Vert_{F}^{2}$ within the
space of positive semi-definite covariances is given by
%
\renewcommand{\theequation}{A.2}
\begin{equation}
\label{minfrobs2} \mathcal{A}^{+} \Biggl(\frac{1}{K}\sum
_{k = 1}^{K} \bolds{\Phi }_{k}^{\prime
}
\mathbf{P}_{k}\bolds{\Phi}_{k} \Biggr).
\end{equation}
\end{prop}

\begin{pf}
By definition of the Frobenius norm,\vspace*{-3pt}
%
\renewcommand{\theequation}{A.3}
\begin{eqnarray}
&&\sum_{k = 1}^{K}\bigl\Vert\mathbf{P}_{k}
- \bolds{\Phi}_{k}\mathbf {C}^{-1}\bolds{\Phi
}_{k}^{\prime}\bigr\Vert_{F}^{2} \nonumber\\
&&\qquad= \sum
_{k = 1}^{K}\operatorname{trace} \bigl\lbrace
\bigl(\mathbf{P}_{k} - \bolds{\Phi}_{k}\mathbf{C}^{-1}
\bolds{\Phi }_{k}^{\prime
} \bigr)^{\prime} \bigl(
\mathbf{P}_{k} - \bolds{\Phi}_{k}\mathbf {C}^{-1}
\bolds {\Phi}_{k}^{\prime} \bigr) \bigr\rbrace
\nonumber
\\
\label{frobnormproof}
&&\qquad= \sum_{k = 1}^{K} \bigl\lbrace
\operatorname{trace} \bigl(\mathbf {P}_{k}^{\prime}
\mathbf{P}_{k} \bigr) - 2\times\operatorname{trace} \bigl( \bolds {
\Phi}_{k}^{\prime}\mathbf{P}_{k}^{\prime}\bolds{
\Phi}_{k}\mathbf {C}^{-1} \bigr) + \operatorname{trace}
\bigl(\mathbf{C}^{-2} \bigr) \bigr\rbrace
\\
\nonumber
&&\qquad= \sum_{k = 1}^{K}
\operatorname{trace} \bigl(\mathbf{P}_{k}^{\prime
}\mathbf
{P}_{k} \bigr)- K\times\operatorname{trace} \Biggl\lbrace \Biggl(
\frac
{1}{K}\sum_{k = 1}^{K}\bolds{
\Phi}_{k}^{\prime}\mathbf{P}_{k}\bolds{\Phi
}_{k} \Biggr)^{2} \Biggr\rbrace
\\
&&\qquad\quad{}+K\times \Biggl\llVert \mathbf{C}^{-1} - \frac{1}{K}\sum
_{k = 1}^{K}\bolds{\Phi }_{k}^{\prime}
\mathbf{P}_{k}\bolds{\Phi}_{k}\Biggr\rrVert
_{F}^{2}.
\nonumber
\end{eqnarray}
It follows from Theorem~2.1 of \citet{Higham} that the minimum of
(\ref
{frobnormproof}) is given by equation (\ref{minfrobs}) in the main
document. In a similar manner, if one substitutes $\mathbf{C}$ for
$\mathbf{C}^{-1}$ in (\ref{frobnormproof}), then we obtain the result
in equation (\ref{minfrobs2}) in the main document.
\end{pf}

\begin{prop}\label{prop2}
Let $\mathbf{S}_{X,1}$ be the MI propagator
matrix and $\mathbf{C}$ be a generic $r \times r$ positive definite
matrix. Then, the value of $\mathbf{C}$ that minimizes $\Vert\mathbf
{Q}_{1} - \mathbf{S}_{X,1}\mathbf{C}\mathbf{S}_{X,1}^{\prime
}\Vert_{F}^{2}$ within the space of positive semi-definite covariances is
given by
%
\begin{equation}
\renewcommand{\theequation}{A.4}
\label{minfrobscor} \mathbf{C}^{*} = \mathcal{A}^{+} \bigl(
\mathbf{S}_{X,1}^{\prime
}\mathbf {Q}_{1}
\mathbf{S}_{X,1} \bigr).
\end{equation}
\end{prop}

\begin{pf}
The proof of Proposition~\ref{prop2} follows
immediately from Proposition~\ref{prop1}. Specifically, let $K = 1$, $\bolds{\Phi
}_{1} = \mathbf{S}_{X,1}$, and $\mathbf{P}_{1} = \mathbf{Q}_{1}$. Then,
apply Proposition~\ref{prop1}. If $\mathbf{S}_{X,1}^{\prime}\mathbf
{Q}_{1}\mathbf
{S}_{X,1}$ is positive definite, then (\ref{minfrobscor}) leads to the
prior specification in \citet{hughes}. \citet{aaronpBayes} show that
$\mathbf{S}_{X,1}^{\prime}\mathbf{Q}_{1}\mathbf{S}_{X,1}$ is positive
definite as long as an intercept is included in the definition of
$\mathbf{X}_{1}$.
\end{pf}

\section{Full conditional distributions}\label{appB}

\renewcommand{\theequation}{B.\arabic{equation}}
The model that we use for multivariate spatio-temporal data is given by
%
\setcounter{equation}{0}
\begin{eqnarray}
&&\quad \mbox{Data model: } Z_{t}^{(\ell)}(A)\vert
\bfbeta_{t}, \bolds{\eta}_{t}, \xi_{t}^{(\ell)}(
\cdot)\nonumber\\
&&\qquad\quad\qquad\qquad\ind \operatorname{Normal} \bigl(\mathbf{x}_{t}^{(\ell)}(A)^{\prime}
\bfbeta_{t} + \mathbf {S}_{X,t}^{(\ell)}(A)^{\prime}
\bolds{\eta}_{t} + \xi_{t}^{(\ell)},
v_{t}^{(\ell)}(A) \bigr);
\nonumber
\\
\nonumber
&&\quad\mbox{Process model 1: } \bolds {\eta }_{t}\vert\bolds{
\eta}_{t-1},\mathbf{M}_{B,t},\mathbf{W}_{t}\sim
\operatorname{Gaussian} (\mathbf{M}_{B,t}\bolds{\eta}_{t-1},
\mathbf {W}_{t} );
\\
\nonumber
&&\quad \mbox{Process model 2: } \bolds {\eta }_{1}\vert
\mathbf{K}_{1} \sim \operatorname{Gaussian} (\mathbf{0},
\mathbf{K}_{1} );
\\
\nonumber
&&\quad\mbox{Process model 3: } \xi _{t}^{(\ell)}(\cdot)
\vert\sigma_{\xi,t}^{2} \ind \operatorname{Normal} \bigl(0,
\sigma_{\xi,t}^{2} \bigr);
\\
\nonumber
&&\quad\mbox{Parameter model 1: } \delta ^{(k)} \sim\operatorname{IG} (
\alpha_{v}, \beta_{v} );
\\
\nonumber
&&\quad \mbox{Parameter model 2: } \bfbeta _{t} \sim \operatorname{Gaussian}
\bigl(\bolds{\mu}_{\beta}, \sigma_{\beta
}^{2}
\mathbf{I}_{p} \bigr);
\\
\nonumber
&&\quad\mbox{Parameter model 3: } \sigma _{\xi,t}^{2}
\sim\operatorname{IG} (\alpha_{\xi}, \beta_{\xi
} );
\\
&&\quad\mbox{Parameter model 4: } \sigma _{K}^{2} \sim
\operatorname{IG} (\alpha_{K}, \beta_{K} );
\nonumber
\\
\eqntext{\displaystyle\ell= 1,\ldots,L, t =
T_{L}^{(\ell)},\ldots,T_{U}^{(\ell)}, k = 1,2, A
\in D_{\mathrm{O},t}^{(\ell)},}
\end{eqnarray}
where $\sigma_{\beta}^{2}>0$, $\alpha_{v}>0$, $\alpha_{\xi}>0$,
$\alpha
_{K}>0$, $\beta_{v}>0$, $\beta_{\xi}>0$, and $\beta_{K}>0$. In
Sections~\ref{analysis} and \ref{secdisc} the prior mean of $\bolds{\mu}_{\beta}$ is set equal
to a
$p$-dimensional zero vector, and the corresponding variance $\sigma
_{\beta}^{2}$ is set equal to $10^{15}$ so that the prior on $\{\bolds
{\beta}_{t}\}$ is vague. In Sections~\ref{analysis} and \ref{secdisc} we also specify $\alpha
_{\xi}$, $\alpha_{K}$, $\beta_{v}$, $\beta_{\xi}$, and $\beta
_{K}$ so
that the prior distributions of $\sigma_{\xi,t}^{2}$ and $\sigma
_{K,t}^{2}$ are vague. Specifically, we let $\alpha_{\xi}=\alpha
_{K}=2$, and $\beta_{v}=\beta_{\xi}=\beta_{K}=1$; here, the $\mathrm{IG}(2,1)$
prior is interpreted as vague since it has infinite variance.

We now specify the full conditional distributions for the process
variables [i.e., $\{\bolds{\eta}_{t}\}$ and $\{\xi_{t}^{(\ell)}(\cdot
)\}$]
and the parameters [i.e., $\{v_{t}^{(\ell)}(\cdot)\}$, $\{\bolds{\beta
}_{t}\}$, $\{\sigma_{\xi,t}^{2}\}$, and $\sigma_{K}^{2}$].

\subsection*{Full conditionals for process variables}
Let the $n_{t}$-dimensional random vectors $ \bz_{t} \equiv (Z_{t}^{(\ell
)}(A): \ell = 1,\ldots,L, A \in D_{\mathrm
{O},t}^{(\ell
)} )^{\prime}$, $ \bolds{\xi}_{t} \equiv (\xi_{t}^{(\ell)}(A):
\ell= 1,\ldots,L, A \in D_{\mathrm{O},t}^{(\ell)} )^{\prime}$,
and the $n_{t}\times p$ matrix $\mathbf{X}_{t} \equiv (\mathbf
{x}_{t}^{(\ell)}(A):\ell= 1,\ldots,L, A \in
D_{\mathrm
{O},t}^{(\ell)} )^{\prime}$; $t = 1,\ldots,T$. Then, we update the
full conditional for $\bolds{\eta}_{1:T} \equiv (\bolds{\eta
}_{t}^{\prime
}: t = 1,\ldots,T )^{\prime}$ at each iteration of the Gibbs
sampler using the Kalman smoother. We accomplish this by performing the
following steps:
\begin{longlist}[3.]
\item[1.] Find the Kalman filter using the shifted measurements $\{
\widetilde{\bz}_{t}: \widetilde{\bz}_{t} = \bz_{t}-\mathbf
{X}_{t}\bolds
{\beta}_{t}-\bolds{\xi}_{t}\}$ [\citet{shumway,cart1994,schnatter94,cressie-wikle-book}]. That is, for $t =
1,\ldots,T$\vspace*{-2pt} compute
\begin{eqnarray*}
&& \mbox{(a) } \bolds{\eta}_{t\vert t}^{[j]} \equiv E
\bigl(\bolds{\eta}_{t}\vert \widetilde{\bz}_{1:t}, \bolds{
\theta}_{t}^{[j]} \bigr),
\\
&& \mbox{(b) } \bolds{\eta}_{t\vert(t-1)}^{[j]} \equiv E \bigl(\bolds{
\eta }_{t}\vert \widetilde{\bz}_{1:(t-1)}, \bolds{
\theta}_{t}^{[j]} \bigr),
\\
&&\mbox{(c) } \mathbf{P}_{t\vert t}^{[j]} \equiv \operatorname{cov}
\bigl(\bolds {\eta }_{t}\vert\widetilde{\bz}_{1:t}, \bolds{
\theta}_{t}^{[j]} \bigr),
\\
&&\mbox{(d) } \mathbf{P}_{t\vert(t-1)}^{[j]} \equiv \operatorname{cov}
\bigl(\bolds {\eta }_{t}\vert\widetilde{\bz}_{1:(t-1)}, \bolds{
\theta}_{t}^{[j]} \bigr),
\end{eqnarray*}
where $\mathbf{P}_{1\vert1}^{[j]} = (\sigma_{K}^{[j]})^{2}\mathbf
{K}^{*}$ and $\bolds{\theta}_{t}^{[j]}$ represents the $j\-$th MCMC draw
of $\bolds{\theta}_{t}$ and $\sigma_{K}^{2}$, respectively.
\item[2.] Sample\vspace*{1pt} $\bolds{\eta}_{T}^{[j+1]} \sim \operatorname{Gaussian} (\bolds
{\eta
}_{T\vert T}^{[j]}, \mathbf{P}_{T\vert T}^{[j]} )$.
\item[3.] For $t = T-1,T-2,\ldots,1$ sample
\[
\bolds{\eta}_{t}^{[j+1]} \sim \operatorname{Gaussian} \bigl(
\bolds{\eta}_{t\vert t}^{[j]} + \mathbf {J}_{t}^{[j]}
\bigl(\bolds{\eta}_{t+1}^{[j]} - \bolds{\eta}_{t+1\vert t}^{[j]}
\bigr), \mathbf{P}_{t\vert t}^{[j]} - \mathbf{J}_{t}^{[j]}
\mathbf {P}_{t+1\vert
t}^{[j]}\bigl(\mathbf{J}_{t}^{[j]}
\bigr)^{\prime} \bigr),
\]
where $\mathbf{J}_{t}^{[j]} \equiv\mathbf{P}_{t\vert t}^{[j]} \mathbf
{M}_{t}^{\prime
}(\mathbf{P}_{t+1\vert t}^{[j]})^{-1}$.
\end{longlist}
Notice that within each MCMC iteraction we need to compute the Kalman
filter and Kalman smoothing equations. This adds more motivation for
reduced rank modeling, that is, if $r$ is large (i.e., if $r$ is close
in value to $n$), this step is not computationally feasible.

The full conditional for the remaining process variable $\{\xi
_{t}^{(\ell)}(\cdot)\}$ can also be computed efficiently [\citet{ravishank}]. The full conditional for $\{\xi_{t}^{(\ell)}(\cdot)\}$ is
given by $\bolds{\xi}_{t} \sim\operatorname{Gaussian} (\bolds{\mu}_{\xi
,t}^{*},\bolds{\Sigma}_{\xi.t}^{*} )$, where {$\bolds{\Sigma}_{\xi
,t}^{*} \equiv (\mathbf{V}_{t}^{-1} + \sigma_{\xi}^{-2}\mathbf
{I}_{N_{t}} )^{-1}$}, $\bolds{\mu}_{\xi,t}^{*} \equiv\bolds{\Sigma
}_{\xi
}^{*}\times\mathbf{V}_{t}^{-1}\times(\bz_{t}-\mathbf{X}_{t}\bolds
{\beta
}_{t}-\mathbf{S}_{t}\bolds{\eta}_{t})$, $\mathbf{V}_{t} \equiv\operatorname{diag} (v_{t}^{(\ell)}(A): \ell= 1,\ldots,L, A \in D_{\mathrm
{O},t}^{(\ell)} )$, and $\mathbf{S}_{t} \equiv
 (\mathbf{S}_{t}^{(\ell)}(A): \ell= 1,\ldots, L,A
\in D_{\mathrm{O},t}^{(\ell)} )^{\prime}$; $t = 1,\ldots,T$.

\subsection*{Full conditionals for the parameters}
Similar to the full
conditional for $\{\xi_{t}^{(\ell)}(\cdot)\}$ [\citet{ravishank}], we
also have the following full conditional for ${\bolds{\beta}_{t}}$:
$\bolds{\beta}_{t} \sim\operatorname{Gaussian} (\bolds{\mu}_{\beta
,t}^{*},\bolds
{\Sigma}_{\beta,t}^{*} )$, where $\bolds{\Sigma}_{\beta,t}^{*}
\equiv
 ( \mathbf{X}_{t}^{\prime}\mathbf{V}_{t}^{-1}\mathbf{X}_{t}+
\sigma
_{\beta}^{-2}\mathbf{I}_{p}  )^{-1}$, and
$\bolds{\mu}_{\beta,t}^{*} \equiv\bolds{\Sigma}_{\beta}^{*}\times
\mathbf
{X}_{t}^{\prime}\mathbf{V}_{t}^{-1}(\bz_{t}-\bolds{\xi}_{t}-\mathbf
{S}_{t}\bolds{\eta}_{t})$; $t = 1,\ldots,T$. The exact form of the full
conditionals for $\sigma_{K}^{2}$ and $\{\sigma_{\xi,t}^{2}\}$ can also
be found in a straightforward manner. It follows that the full
conditionals for $\sigma_{K}^{2}$ and $\sigma_{\xi,t}^{2}$ are $\operatorname{IG}(\operatorname{Tr}/2
+ 2, 1 + \bolds{\eta}_{1}^{\prime}\mathbf{K}_{1}^{*-1}\bolds{\eta}_{1}/2 +
\sum_{t=2}^{T}(\bolds{\eta}_{t}-\mathbf{M}_{t}\bolds{\eta
}_{t-1})^{\prime
}\mathbf{W}_{t}^{*-1}(\bolds{\eta}_{t}-\mathbf{M}_{t}\bolds{\eta}_{t-1})/2)$
and IG($n/2 + 2$, $1 + \bolds{\xi}_{t}^{\prime}\bolds{\xi}_{t}/2$) (for
$t =
1,\ldots,T$), respectively.

{Imputation variances for QWIs are not currently available for each
county/\break quarter/industry/gender combination, which is the
multivariate spatio-\break temporal support of the data in Section~\ref{secmotiv}. Thus, we specify a prior distribution for $\{v_{t}^{(\ell
)}(A)\}$ that capitalizes on the available information, namely,
imputation variances defined for QWIs given at each
county/quarter/industry combination. Denote these imputation
variances with $\widetilde{v}_{t}^{(m)}(\cdot)$, where $m = 1,\ldots
,20$ and $t = 1,\ldots,92$. This leads us to our prior for $\{
v_{t}^{(\ell)}(A)\}$ given by
\[
v_{t}^{(\ell)}(A) = \cases{ \widetilde{v}_{t}^{(\ell)}(A)
\delta^{(1)}/\operatorname{exp}\bigl\{2Z_{t}^{(\ell)}(A)
\bigr\}, \vspace*{2pt}\cr\qquad \mbox{if } \ell=1,\ldots,20, \vspace*{2pt}
\cr
\widetilde{v}_{t}^{(\ell-20)}(A)\delta^{(2)}/\operatorname{exp}
\bigl\{2Z_{t}^{(\ell)}(A)\bigr\}, \vspace*{2pt}\cr
\qquad\mbox{if } \ell=21,
\ldots,40; t = 1,\ldots,92, A \in D_{\mathrm{O},t}^{(\ell)}, }
\]
where $\delta^{(k)}>0$ for $k = 1,2$, and we let $\ell= 1,\ldots,20$
indicate men in each of the 20 industries and $\ell= 21,\ldots,40$
indicate women in each of the 20 industries, respectively. We divide by
$\operatorname{exp}\{2Z_{t}^{(\ell)}(A)\}$ to transform
$\widetilde{v}_{t}^{(\ell)}$ to the log-scale; specifically, we use the
delta method [see \citet{delta}, among others] to transform the
variances to the log-scale. Thus, our model for the variances $\{
v_{t}^{(\ell)}(A)\}$ is a simple reweighting (by weights in $\{\delta
^{(k)}\}$) of the imputation variances (on the log-scale) obtained from
the LEHD program. We note that our predictions are relatively robust to
this specification.}

In the empirical study in Sections~\ref{s51} and~\ref{s52}, we use the known value
of $v_{t}^{(\ell)}(A)$ and, hence, no distribution was placed on
$\delta
^{(1)}$ and $\delta^{(2)}$. In many cases this is reasonable since the
statistical agency provides values for $v_{t}^{(\ell)}(A)$. In
Section~\ref{secmassive} we let $\delta^{(k)} \sim\operatorname{IG}(1,2)$;
$k = 1,2$. Now, let $\ell= 1,\ldots,20$ indicate the spatial fields
corresponding to each of the 20 industries for men and $\ell=
21,\ldots
,40$ indicate the spatial fields corresponding to each of the 20
industries for women. The full conditionals for $\delta^{(1)}$ and
$\delta^{(2)}$ are $\operatorname{IG}(M/2 + 2$, $1 + \sum_{\ell= 1}^{20}\sum_{t =
1}^{92}\sum_{A \in D_{\mathrm{O},t}^{(\ell)}} (Z_{t}^{(\ell
)}(A)-\mathbf
{x}_{t}^{(\ell)}(A)^{\prime}\bfbeta_{t} - \mathbf{S}_{X,t}^{(\ell
)}(A)^{\prime}\bolds{\eta}_{t} - \xi_{t}^{(\ell)})^{2}/2\widetilde
{v}_{t}^{(\ell)}(A)$) and $\operatorname{IG}(F/2 + 2$, $1 + \sum_{\ell=
21}^{40}\sum_{t = 1}^{92}\sum_{A \in D_{\mathrm{O},t}^{(\ell)}} (Z_{t}^{(\ell
)}(A)-\mathbf{x}_{t}^{(\ell)}(A)^{\prime}\bfbeta_{t} -\mathbf
{S}_{X,t}^{(\ell)}(A)^{\prime}\bolds{\eta}_{t} - \xi_{t}^{(\ell
)})^{2}/2\widetilde{v}_{t}^{(\ell)}(A)$), where $M \equiv\sum_{\ell=
1}^{20}\sum_{t = 1}^{92}n_{t}^{(\ell)}$ and\vspace*{1pt} $F \equiv\sum_{\ell=
21}^{40}\sum_{t = 1}^{92}n_{t}^{(\ell)}$.

In some settings, survey error variances are not provided. The case of
unknown survey variance leads to interesting and difficult modeling
questions. In particular, when var($\epsilon_{t}^{(\ell)}$) =
$v_{t}^{(\ell)}(\cdot)$
is unknown, there may be issues with identifiability between $\sigma
_{\xi,t}^{2}$ and $v_{t}^{(\ell)}(\cdot)$ when $v_{t}^{(\ell
)}(\cdot)$
is roughly constant across variables and locations [see \citet
{bradleyTEST}, for a discussion]. To avoid this issue of
identifiability, one might combine $\xi(\cdot)$ and $\epsilon
_{t}^{(\ell
)}(\cdot)$, and then estimate the sums $\xi(\cdot) + \epsilon
_{t}^{(\ell
)}(\cdot)$ and $v_{t}^{(\ell)}(\cdot) + \sigma_{\xi,t}^{2}$,
respectively. { In the environmental context, o}thers have addressed
this identifiability problem by avoiding the use of likelihoods and
adopting a moment-based approach to estimate $v_{t}^{(\ell)}(\cdot)$;
specifically, {see} \citet{kang-cressie-shi-2010} and \citet
{katzfuss2012} for the definition of a variogram-extrapolation
technique to estimate $v_{t}^{(\ell)}(\cdot)$ and \citet{kang-cressie-shi-2010}
for a method of moments estimator.
\end{appendix}

\section*{Acknowledgments}
We thank the Editor, Associate Editor, and two anonymous referees 
for providing valuable comments that strengthened this\break manuscript.

\printaddresses
\end{document}